\documentclass[twocolumn]{aastex7}

\usepackage{array}
\usepackage{amsmath}	
\usepackage{graphicx}	
\usepackage{caption}
\usepackage{gensymb}
\usepackage{longtable}
\usepackage{subcaption}
\usepackage{placeins}
\usepackage{afterpage}
\usepackage[normalem]{ulem}
\useunder{\uline}{\ul}{}

\newcommand{\NRAO}{\affiliation{National Radio Astronomy Observatory, 520 Edgemont Road, Charlottesville, VA 22903, USA}}

\begin{document}
 
\title{The VLA Frontier Fields Survey: A 6\,GHz High-resolution Radio Survey of  Abell\,2744}

\author[0009-0007-7413-5222]{Esteban A. Orozco}
\altaffiliation{Instituto de Radioastronomía y Astrofísica (IRyA)}
\affiliation{Instituto de Radioastronomía y Astrofísica, Universidad Nacional Autónoma de México, Antigua Carretera a Pátzcuaro \# 8701, Ex-Hda. San José de la Huerta, Morelia, Michoacán, México C.P. 58089}
\affiliation{Observatorio Astronómico de Quito, Escuela Politécnica Nacional, Interior del Parque La Alameda, 170136, Quito, Ecuador}
\email[show]{e.orozco@irya.unam.mx}  

\author[0000-0002-2640-5917]{Eric F. Jim\'enez-Andrade} 
\affiliation{Instituto de Radioastronomía y Astrofísica, Universidad Nacional Autónoma de México, Antigua Carretera a Pátzcuaro \# 8701, Ex-Hda. San José de la Huerta, Morelia, Michoacán, México C.P. 58089}
\NRAO 
\email{e.jimenez@irya.unam.mx}

\author[0000-0001-7089-7325]{Eric J. Murphy} \NRAO \email{emurphy@nrao.edu}

\author[0000-0003-3037-257X]{Ian Smail}
\affiliation{Centre for Extragalactic Astronomy, Department of Physics, Durham University, South Road, Durham DH1 3LE, UK}
\email{ian.smail@durham.ac.uk}

\author[0000-0003-3168-5922]{Emmanuel Momjian}
\affiliation{National Radio Astronomy Observatory, P.O Box O, Socorro, NM 87801, USA}
\email{emomjian@nrao.edu}

\author[0000-0001-6864-5057]{Ian Heywood}
\affiliation{Astrophysics, Department of Physics, University of Oxford, Keble Road, Oxford, OX1 3RH, UK}
\email{ian.heywood@physics.ox.ac.uk}

\author{Miguel A. Vega}
\affiliation{Instituto de Radioastronomía y Astrofísica, Universidad Nacional Autónoma de México, Antigua Carretera a Pátzcuaro \# 8701, Ex-Hda. San José de la Huerta, Morelia, Michoacán, México C.P. 58089}
\email{m.vega@irya.unam.mx}

\author[0000-0002-4781-9078]{Christa DeCoursey}
\affiliation{ Steward Observatory, University of Arizona, 933 N. Cherry Ave, Tucson, AZ 85721, USA}
\email{cndecoursey@arizona.edu}
 
\begin{abstract}

We present 6\,GHz radio continuum observations of the galaxy cluster Abell\,2744 ($z = 0.307$) obtained with the \textit{Karl G. Jansky Very Large Array} (VLA) as part of the VLA Frontier Fields program, the goal of which is to explore the radio continuum emission from high-redshift galaxies that are magnified by foreground massive galaxy clusters. With an rms noise of $\approx 1 \mu$Jy beam$^{-1}$, in the image plane, and sub-arcsec angular resolution ($\theta_{1/2}=0\farcs 82$), this is the deepest and most detailed radio image of Abell\,2744 ever obtained. A total of 93 sources is detected with a peak signal-to-noise ratio $\geq5$, of which 46 have optical/near-infrared (IR) counterparts with available redshift, magnification ($\mu$), and stellar mass (${M}_*$) estimates.  
The radio sources are distributed over a redshift of 0.15 to 3.55, with a median redshift value of $\Bar{z} = 0.93^{+1.48}_{-0.63}$ and with a range mass from $5.5\times 10^{9} \,\rm{M}_{\odot}$ to $1.3\times 10^{11} \,\rm{M}_{\odot}$. A comparison between the radio-based star formation rates (SFRs) and those derived from ultraviolet-to-near IR data reveals that radio SFRs are typically an order of magnitude higher. This discrepancy is likely a result of strong dust obscuration affecting the UV-to-NIR tracers. 
We  look for radio counterparts of the so-called ``Little Red Dots (LRDs)'' galaxies at $z\approx6$ seen behind Abell\,2744, but find no significant detections. After stacking, we derive a 3$\sigma$ upper limit to the 6\,GHz radio luminosity of LRDs of $4.1\times 10^{39}\,\rm erg\,s^{-1}$. Finally, we present a sample of 22 moderately/strongly lensed galaxies ($\mu \gtrsim 2$) in the VLA Frontier Fields survey, which provides a zoomed view of the star formation processes within main sequence  galaxies at $z\approx 1-2$.

\end{abstract}

\keywords{\uat{Radio source catalogs}{1356} --- \uat{Star formation}{1569} --- \uat{Gravitational lensing}{670} --- \uat{Very Large Array}{1766} --- \uat{Abell clusters}{9}}

\section{Introduction} 

The Hubble Frontier Fields (HFF) program originated as a multi-cycle observing campaign using the Hubble (HST) and Spitzer Space Telescopes targeting six strong-lensing galaxy clusters, and six parallel blank fields \citep{lotz2017frontier}. The Frontier Fields take advantage of the gravitational lensing effect provided by the massive clusters, allowing us to characterize the emission from galaxies that are intrinsically faint or lie at high redshifts. A key goal of the Frontier Fields project is to collect a substantial sample of these sources and gain insight into the star-formation processes in the early universe, through measurements of stellar mass, star-formation rates (SFRs), and the structure of high-redshift galaxies. 

The six galaxy clusters of the Frontier Fields program have been observed at different wavelengths with multiple ground and space observatories, including observations with \textit{Herschel} Space Telescope \citep{rawle2016complete}, Atacama Large Millimeter Array \citep[ALMA;][]{gonzales2017, Laporte2017},  and \textit{Chandra} X-ray observatory \citep{van2016discovery, rahaman2021x}. More recently, \cite{heywood2021vla} presented  3 and 6\,GHz observations taken with NRAO's Jansky Very Large Array (VLA) of three HFF clusters (MACSJ0416.1-2403, MACSJ0717.5+3745, and MACSJ1149.5+2223), as part of the VLA Frontier Field project, with a $\sim 1\mu$Jy beam$^{-1}$ sensitivity and a sub-arcsecond resolution ($\sim 2$\,kpc at $z \approx 3$). 
Deep observations at 3 and 6 GHz primarily trace star-forming galaxies (SFGs) at $0<z\lesssim 4$, which emit in the radio band due to synchrotron radiation from electrons accelerated in supernova remnants and free-free continuum emission from hot, ionized H{\sc ii} regions. The link between radio emission and the SFR can be established using the empirically derived far-infrared–radio correlation (FIRRC), which enables characterization of star formation in high-redshift SFGs without the effects of dust obscuration. 
The FIRRC is thought to originate from the star formation process in galaxies, as most massive stars ($\geq 8\,\rm{M}_{\odot}$) radiate mainly at ultraviolet (UV) wavelengths, and a fraction of the UV photons are absorbed and remitted in the infrared (IR) range due to thermal dust emission. After several Myrs, these young massive stars explode as supernovae (SNe), accelerating cosmic rays into the magnetic field of their host galaxy and resulting in diffuse synchrotron emission \citep{helou1985thermal, condon1992radio, murphy2006effect, murphy2008connecting, murphy2009far, magnelli2015far, algera2020alma}. In essence, massive stars provide a common origin for the far infrared and synchrotron emission. With respect to others SFR tracers, such as the UV and H$\alpha$ luminosities, the radio emission is unaffected by dust extinction and characterizes efficiently the emission from dust-obscured star-formation \citep[e.g., ][]{chapman2004population}

In the initial data paper by \cite{heywood2021vla}, 1966 compact radio components at signal-to-noise ratio (SNR $>5$) are reported across three fields (MACS\,J0416.1-2403, MACS\,J0717.5+3745, and MACS\,J1149.5+2223) at 3\,GHz and 6\,GHz, of which 169 are reported from the narrower 6\,GHz maps; 1262 have spectroscopic redshifts ($z_{\rm spec}$) and 55 have photometric redshifts ($z_{\rm phot}$), with a median $\bar{z}_{\rm phot}=0.88$. Using the highest
resolution ($\sim0\farcs3$) C-band images, \cite{heywood2021vla} report a median angular radio size of $0\farcs27\pm0\farcs25$.
 They detected a total of 13 moderately lensed ($2.1 < \mu < 6.5$) sources, including a radio source that has a demagniﬁed peak brightness of $0.9\,\mu$Jy beam$^{-1}$, making it a candidate for the faintest extragalactic radio source ever detected \citep{heywood2021vla}. Using the 3 and 6\,GHz images, \cite{jimenez2021vla} derived the median 3\,GHz radio sizes of $R_{\rm{eff}} = 1.3\pm0.3$\,kpc for a sample of 98 star-forming galaxies spanning $0.3 \lesssim z \lesssim 3$, with a median stellar mass of $\log (M_\star/\rm{M}_{\odot}) \approx 10.4$. These measurements were compared with the UV/optical sizes derived from {\it HST} ACS/WFC3 imaging, while the radio continuum traces the bulk of massive star formation, and the optical emission predominantly traces the stellar disk of galaxies. \citet{jimenez2021vla} found that the radio size decreases as the SFR increases, and that the optical size is a factor $\approx2-3$ larger than that measured in the radio ---hinting at centrally enhanced star formation activity in these radio-selected SFGs, similar to the dust obscured star formation in ALMA-detected submillimeter galaxies \citep[e.g., ][]{gillman2024structure}. To further contribute to the characterization of radio continuum emission from high-redshift galaxies, here we present the 6\,GHz image and associated radio source catalog of Abell\,2744.

Abell\,2744 \citep[hereafter A\,2744; a.k.a. AC118;][]{olowin1988x} is a massive X-ray galaxy cluster at $z = 0.3072$. Its strong gravitational lensing effect \citep[e.g.,][]{smail1991statistically} made it one of the six massive clusters selected for the \textit{HST} Frontier Fields project \citep{lotz2017frontier}. A\,2744 is located at RA (J2000) = 00h 14m 20.03s and DEC (J2000) = $-$30h 23m 17.80s, with a virial mass of $7.4 \times 10^{14} \rm{M}_{\odot}$ \citep{moretti2022observing}. A\,2744 has been extensively observed at multiple wavelengths \citep[e.g.,][]{moretti2022observing, fujimoto2023dualz, wang2023uncover}, because it is visible from both the northern and southern observatories. A\,2744 is also the third strongest lensing galaxy cluster among the six Frontier Fields, which increases the likelihood of identifying strongly lensed systems at high redshifts.

Several observational programs have targeted A\,2744 at multiple radio and millimeter (mm) wavelengths. The sensitivities, observed frequencies, and angular resolutions of these programs are presented in Table \ref{tab:observational_programs}. We highlight \citet{pearce2017vla} who, in particular, observed A\,2744 with the VLA using the L-band (1-2\,GHz) and  S-band (2-4\,GHz) receivers in the DnC-, CnB-, and BnA-array configurations, leading to  $\sim 4-10\,\mu$Jy\,beam$^{-1}$ sensitivities and $\approx 1\farcs5 - 4\farcs0$ angular resolutions.  
Our 6\,GHz image with an angular resolution of $0\farcs82 \times 0\farcs82$ and depth of $\sim 1\,\mu$Jy\,beam$^{-1}$ is, therefore, the deepest and sharpest radio/sub-mm image yet obtained of this field. \\

\begin{table*}[]
\centering
\footnotesize
\begin{tabular}{cccccc}
\hline
\hline
Reference       &  Observation Date & Instrument & Frequency [GHz]       & $\sigma$[$\mu$Jy\,beam$^{-1}$] & Angular resolution \\ \hline
 \cite{gonzales2017}  & 29-Jun.-2014/31-Dec.-2014   & ALMA       & 263.14         & 55    & $\approx 2\farcs2 \times 2\farcs1$           \\
 \cite{pearce2017vla}          & 22-Sep.-2014/02-Jun.2015      & VLA        &  1-2         & 10    & $4\farcs15\times 2\farcs83$            \\
 \cite{pearce2017vla}          & 20-Sep.-2014/22-May.2015      & VLA        &  2-4         & 4.1    & $1\farcs65\times 1\farcs40$            \\
  \cite{paul2019low} &  24-Nov.-2006/24-Nov.-2006      & GMRT$^{*}$       & 0.235 \& 0.610 & $10^4$   & $3\farcs5 \times 3\farcs5$            \\
 \cite{fujimoto2023dualz}        & 04-Oct.-2022/04-Oct.-2022      & ALMA        & 230         & 32.7  & $1\farcs5 \times 1\farcs5$           \\
 This work  & 04-Dec.2016/29-Apr.2022 & VLA & 4-8 & $\sim 1$ & $0\farcs82 \times 0\farcs82$ \\ \hline
\end{tabular}
\caption{Summary of radio programs that have observed A\,2744 at multiple wavelengths, employing different angular resolutions and sensitivities. Column 1: Key publication describing the dataset. Column 2: Observation date range. Column 3: Observatory and/or instrument used to acquire the data. $^{*}$GMRT: \textit{Giant Metrewave Radio Telescope}. Column 4: Observing frequency. Column 5: Root-mean-square (rms) noise level ($\sigma$) of the maps. Column 6: Achieved angular resolution of the radio observations of A\,2744.}
\label{tab:observational_programs}
\end{table*}

This paper describes the observations, data reduction, and the production and validation of the 6\,GHz  radio data products in the A\,2744 field. 
An overview of the VLA radio observations of A\,2744, the imaging methods, and the data reduction process is described in Section \ref{section: Data}. The source extraction method and the radio sizes of our 6\,GHz sample are presented in Section \ref{section:Source extraction}. The counterpart association is discussed in Section \ref{section: Counterpart asociation}. Section \ref{sec: Results_section} presents the properties of our source catalog, including magnification, redshift, and stellar mass distributions. It also describes the methods used for active galactic nuclei (AGN) identification, the derivation and comparison of SFRs based on radio and $u$-band data, a comparison of the specific star formation rates (sSFRs) with those from other studies focused on magnified galaxies, and the procedure to search for the puzzling population of ``Little Red Dots'' (LRDs) in our map. We summarize our results in Section \ref{sec: Results_section}. The assumed cosmological model throughout this paper is $\Lambda$-CDM with H$_0 = 70\,$km s$^{-1}$ Mpc$^{-1}$, $\Omega_{\rm{M}} = 0.3$, and $\Omega_{\Lambda} = 0.7$.

\section{Data}\label{section: Data}
\subsection{VLA Data, Calibration, and Imaging}
The data were obtained from the VLA projects 16B-319 (PI: E. Murphy) and 22A-017 (PI: E. Jimenez-Andrade), as part of the VLA Frontier Fields Survey \citep{heywood2021vla}. These observational campaigns included observations in the A and C configurations using C-band receivers (4–8\,GHz), which provide a primary beam radius of $\theta_{\textrm{PB}}=3\farcm75$, resulting in a high spatial resolution while capturing diffuse and extended radio emission.
The respective integration time on-source was 3.75 hours for the 22A-017 project, with 3 scheduling blocks in the C configuration; while for the 16B-319 project, the time on-source was 7 hours, with 7 scheduling blocks in the C configuration and 1 scheduling block in the A configuration. 
Each of the 11 scheduling blocks (SB) was processed with the NRAO \texttt{CASA} pipeline version  2022.2.0.64. The pipeline performed automated flagging of data affected by antenna shadowing, zero visibility amplitudes, and the initial integrations during the antenna slewing. Additionally, a first pass of radio frequency interference (RFI) excision was applied to both calibrator and target scans. Following the execution of the pipeline, spectral windows (SPWs) with anomalously high amplitudes and/or RFI were identified and flagged.

To produce the 6\,GHz continuum image of A\,2744, the \texttt{tclean} task in \texttt{CASA} was used. The image size was set to 5808 pixels or $871\farcs2$ to contain a circularized synthesized beam of $\theta_{1/2}=0\farcs82$, with a pixel size of $0\farcs15$. The imaging process was configured with a maximum of 20,000 iterations and a stopping threshold of $3\,\mu\rm  Jy\,beam^{-1}$, corresponding to three times the expected noise level.
A Briggs weighting scheme with a robustness parameter of 0.5 was chosen to balance resolution and noise suppression. We adopted the continuum imaging mode (\texttt{specmode = mfs}) including a spectral polynomial fit with two terms (\texttt{nterms = 2}) to optimize wide-band imaging. The multi-term, multi-frequency synthesis deconvolver (\texttt{deconvolver = mtmfs}) was used, as it is the recommended approach for wide-band, wide-field imaging of sources with varying physical scales.
Deconvolution was performed using multiscale cleaning with the  wide-field gridding algorithm (\texttt{gridder = widefield}) and 64 \texttt{w-projection} planes.
A primary beam limit of \texttt{pblimit = 0.05} was applied to maximize the inclusion of sources without compromising flux reliability.

\begin{figure*}
    \centering
    \includegraphics[width=1\linewidth]{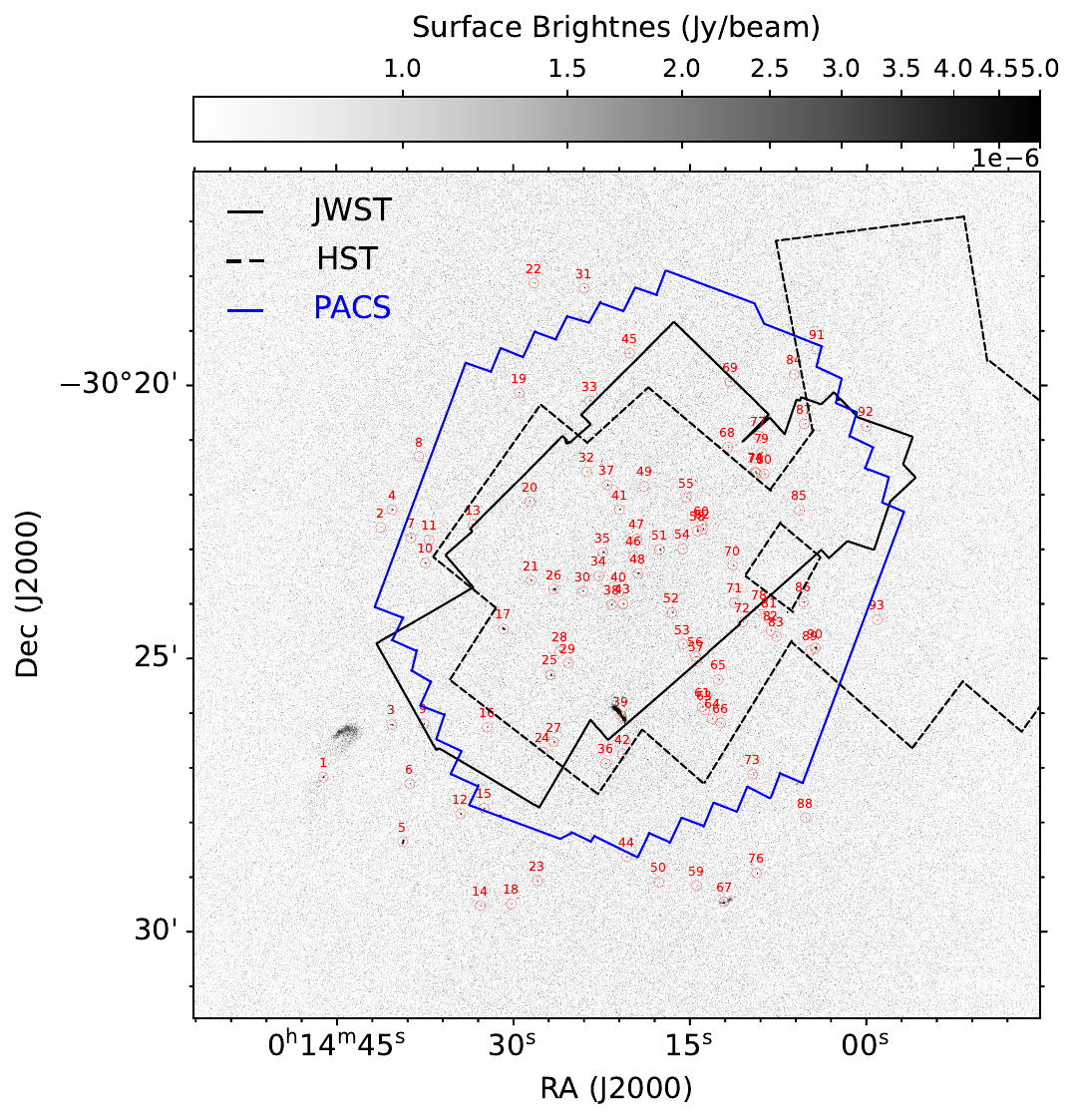}
    \caption{Our 6\,GHz image of A\,2744, before PB correction. The white contour outlines the \textit{JWST} footprint from the UNCOVER survey, while the pink dashed contour marks the \textit{HST} footprint from the same survey. The green contour marks the PACS area from the \textit{Herschel} observations. Red circles and numbers denote the positions and short IDs (see Table\,\ref{tab:VLA sources}) of the radio sources detected in this study.}
    \label{fig:VLA_ALMA_footprint}
\end{figure*}

\begin{figure}
    \centering
    \includegraphics[width=1\linewidth]{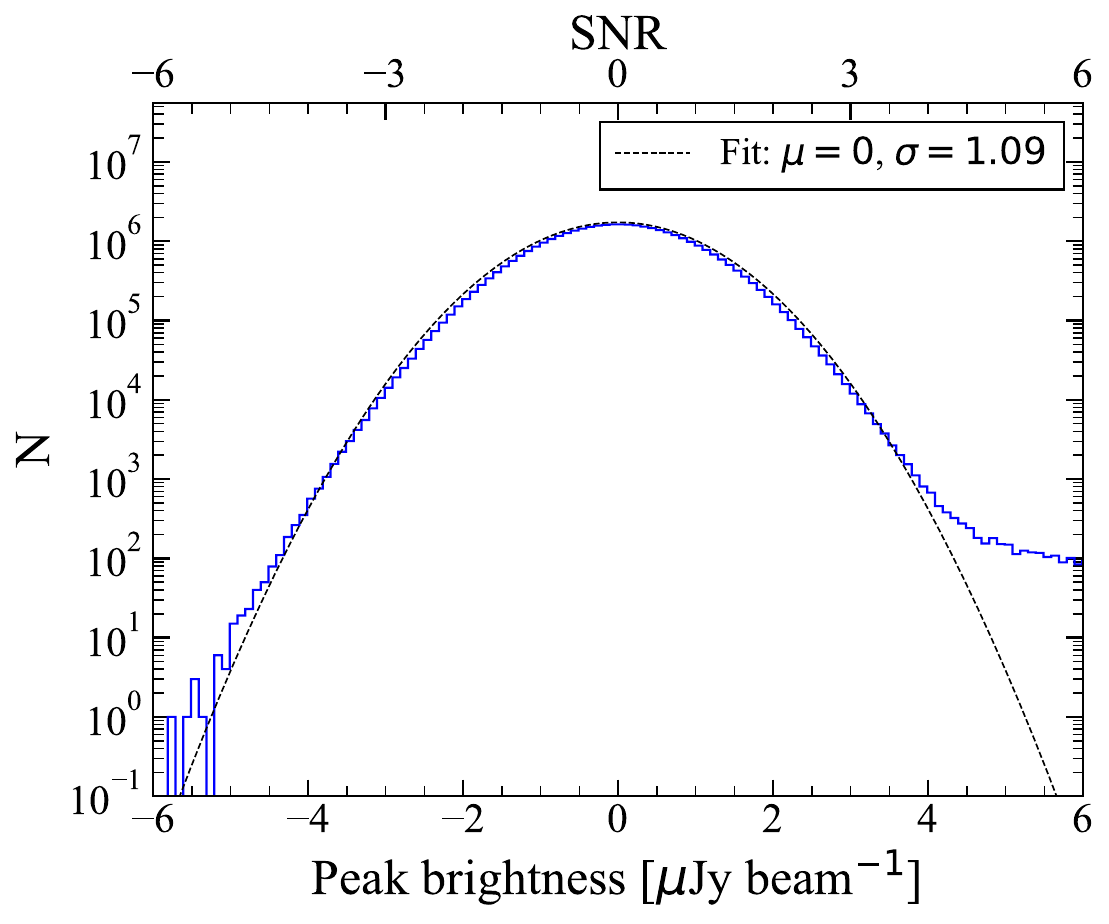}
    \caption{Pixel brightness distribution from the 6\,GHz A\,2744 map -- prior to primary beam correction. The bin width is 0.1\,$\mu$Jy\,beam$^{-1}$. The dashed line is a Gaussian fit with $\sigma = 1.09\,\mu$Jy\,beam$^{-1}$ (i.e., the observed rms noise of the image prior to primary beam correction).}
    \label{fig:pixel_brigthness_distribution}
\end{figure}

With this set of parameters, we generated an image with a native resolution of $\theta_{1/2}=0\farcs84 \times 0\farcs 23$ at a position angle of $5\degree$, where $\theta_{1/2}$ refers to the full width at half maximum (FWHM) along the major/minor axis of the synthesized beam, respectively. Later, we generated a new version of the map with a circularized synthesized beam of $\theta_{1/2}=0\farcs82$ (Figure \ref{fig:VLA_ALMA_footprint}).
The pixel brightness distribution of the map, before primary beam correction, is accurately described by a Gaussian function with a standard deviation ($\sigma = 1.09\,\mu$Jy beam$^{-1}$; see Figure\,\ref{fig:pixel_brigthness_distribution}). The few deviations from a Gaussian model in the negative end of the histogram are mainly related to the presence of imaging artifacts or spurious sources (see Section \ref{section:Source extraction}), while the positive deviations are associated with the radio sources detected in our 6\,GHz VLA image.

\subsection{HST + JWST}
\begin{figure}
    \centering
    \includegraphics[width=1\linewidth]{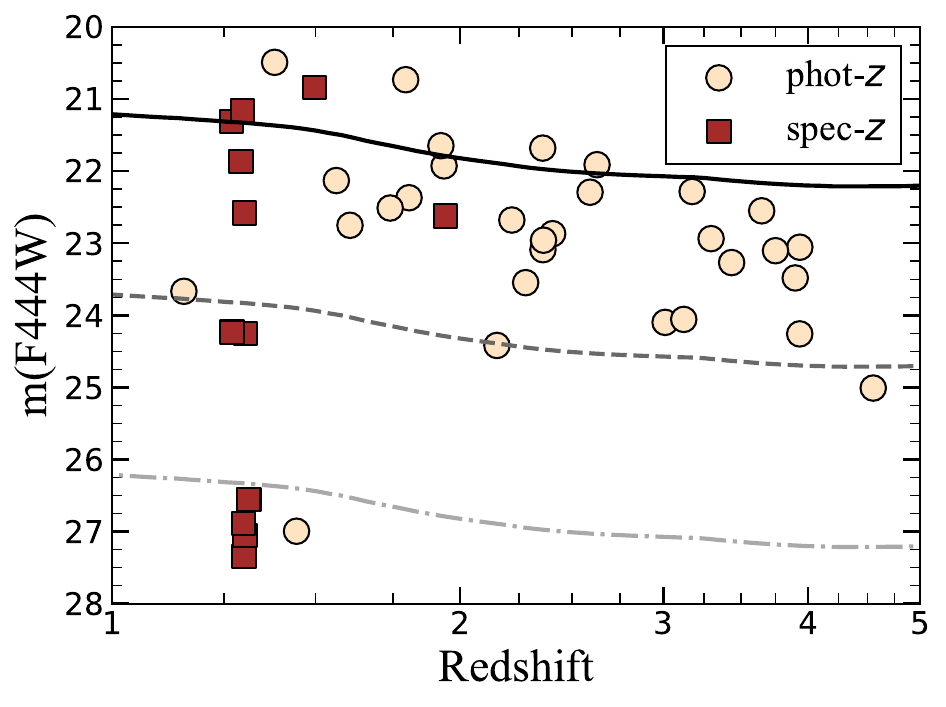}
    \caption{F444W apparent AB magnitude vs. redshift scaled as log($1+z$) for our 6\,GHz sample. Shapes and colors indicate the type of redshift. Red squares correspond to sources with spectroscopic redshifts, while yellow circles indicate those with  photometric redshifts. The black solid line shows the F444W magnitude of a stellar population modeled with \citet{bruzual2003stellar} and a stellar mass of $10^{11}\,\mathrm{M}{\odot}$. The gray dashed line corresponds to a stellar mass of $10^{10}\,\mathrm{M}{\odot}$, and the gray dot–dashed line represents a stellar mass of $10^{9}\,\mathrm{M}_{\odot}$.}
    \label{fig:F444Wvsz}
\end{figure}

The optical and near infrared (NIR) observations come from the ``UNCOVER Photometric Catalog'' \citep{wang2023uncover} that  includes all the available JWST/NIRCam imaging and \textit{HST} ACS/WFC3 deep observations of A\,2744 (see Table \ref{tab:JWST_HST_programs}).
\begin{table*}[]
\centering
\begin{tabular}{cccc}
\hline
\hline
\multicolumn{4}{c}{James Webb Space Telescope (JWST)} \\ \hline
Program &
  \begin{tabular}[c]{@{}c@{}}PI.\\ Reference\end{tabular} &
  Instrument &
  Filters \\ \hline
JWST -GO-2561 &
  \begin{tabular}[c]{@{}c@{}}R. Bezanson\\ \cite{bezanson2024jwst}\end{tabular} &
  NIRCam &
  \begin{tabular}[c]{@{}c@{}}F115W, F150W, F200W, \\ F277W, F356W, F410W, and F444W\end{tabular} \\
JWST -DD-ERS-1324 &
  \begin{tabular}[c]{@{}c@{}}T. Treu\\ \cite{treu2022glass}\end{tabular} &
  NIRCam &
  \begin{tabular}[c]{@{}c@{}}F090W, F115W, F150W,\\ F200W, F277W, F356W, and F444W\end{tabular} \\
JWST-DD-276 &
  \begin{tabular}[c]{@{}c@{}}W. Chen\\ \cite{chen2022imaging}\end{tabular} &
  NIRCam &
  \begin{tabular}[c]{@{}c@{}}F115W, F150W, F200W, \\ F277W, F356W, and F444W\end{tabular} \\ \hline
\multicolumn{4}{c}{Hubble Space Telescope (HST)} \\ \hline
Program &
  \begin{tabular}[c]{@{}c@{}}PI.\\ Reference\end{tabular} &
  Instrument &
  Filters \\ \hline
HST -GO-11689 &
  \begin{tabular}[c]{@{}c@{}}R. Dupke\\ \cite{lotz2017frontier}\end{tabular} &
  ACS/WFC &
  F435W, F606W, and F814W \\
HST -GO 13386 &
  \begin{tabular}[c]{@{}c@{}}S. Rodney\\ \cite{lotz2017frontier}\end{tabular} &
  ACS/WFC &
  F105W, F160W, F606W, and F814W \\
HST -DD-1395 &
  \begin{tabular}[c]{@{}c@{}}J. Lotz\\ \cite{lotz2017frontier}\end{tabular} &
  WFC3/IR &
  F105W, F125W, and F140W \\
HST -GO-15177 &
  \begin{tabular}[c]{@{}c@{}}C. Steinhardt \& L. Jauzac\\ \cite{steinhardt2020buffalo}\end{tabular} &
   \begin{tabular}[c]{@{}c@{}}ACS/WFC\\  WFC3/IR\end{tabular} &
  \begin{tabular}[c]{@{}c@{}}F606W, F814W, F105W, \\ F125W, and F160W\end{tabular} \\ \hline
\end{tabular}
\caption{All the available \textit{JWST}/NIRCam imaging and HST ACS/WFC3 deep observations of A\,2744 that compose the ``UNCOVER Photometric Catalog'' \citep[UNCOVER;][]{2024ApJS..270....7W}. Column 1: Observation name/code. Column 2: Proposal ID (PI.) and reference reporting the dataset and its corresponding observational characteristics. Column 3: Instrument used to acquire the data. Column 4: Filters used in the observation. In total, the “UNCOVER Survey” employs 8 \textit{JWST} filters and 7 \textit{HST} filters extending the sky coverage around A\,2744, allowing the inclusion of nearby cluster sub-structures.}
\label{tab:JWST_HST_programs}
\end{table*}
UNCOVER provides an ultra-deep, noise-equalized detection image using the F277W+F356W+F444W filters covering $\approx0.4-4.4\, \mu m$. The aperture photometry was measured on images whose PSF was matched to the resolution of the F444W image. The F444W filter was adopted as the PSF target because it is the longest-wavelength NIRCam band. This process was essential to  preserve a consistent, common resolution across bands, while probing the reddest rest-frame stellar light \citep[$\approx1–2\, \mu m$ at $z>1$;][]{wang2023uncover}. 
The typical $5\sigma$ detection limit of the F444W is $\sim 29.2$ AB \citep{wang2023uncover}. As observed in Figure \ref{fig:F444Wvsz}, the majority of JWST counterparts of our 6\,GHz radio sources are bright in the F444W image (median AB magnitude of $\sim 25$), i.e., roughly 3 magnitudes brighter than  the $5\sigma$ limits reported by \citet{wang2023uncover}. This indicates the contamination fraction when matching $\mu$Jy radio sources to NIR sources is small. These properties make the {\it JWST} counterparts excellent targets for multi-wavelength follow-up to clarify their nature.
\subsection{Herschel}
The Herschel Space Observatory data are composed of PACS and SPIRE observations. The PACS observations were performed at 100 and 160\,$\mu$m bands. These observations consist of two orthogonal scan maps with a total observing time of 4.4\,h. The observations were part of the PACS Evolutionary Probe \citep[PEP; ][]{lutz2011pacs} and H-ATLAS survey \citep[PI: S. Eales; ][]{eales2010herschel}. The observations were completed in parallel (PACS + SPIRE) mode and achieved a $5\sigma$ depth at $100\,\mu m$ of 120\,mJy. The PACS images extend to a radius of 4-5\,arcmin from the cluster core \citep{rawle2016complete}, as is observed in Figure\,\ref{fig:VLA_ALMA_footprint}. The beam sizes have FWHM=$5\farcs2$, $7\farcs7$, and $12\farcs0$ at 70, 100, and 160\,$\mu$m, respectively. The SPIRE observations were carried out at 250, 350, and $500\,\mu$m, extending to 10 arcmin from the cluster core. The corresponding beam sizes are $18\farcs0$, $25\farcs0$, and $36\farcs0$, respectively, and the observations reach $5\,\sigma$ limits of 28, 32, and 33\,mJy \citep{nguyen2010hermes}.

\section{Source Extraction}\label{section:Source extraction}
Since a homogeneous rms noise across images simplifies the source extraction procedure, we adopted the 6\,GHz map of A\,2744 before primary beam correction to obtain our radio source catalog. To this end, we used the Python Blob Detector and Source Finder \citep[\texttt{PyBDSF};][]{mohan2015pybdsf} using the parameters listed in Table \ref{tab:parameters_PyBDSF}. We imposed an SNR threshold of 5 to detect peaks of emission and an SNR threshold of 3 to identify islands of emission, resulting in 104 extracted islands. 
In some cases, \texttt{PyBDSF} identifies extended emission from nearby bright radio sources as separate islands. To avoid cataloging these artifacts as individual sources, we performed a visual inspection. We found 16 such islands originating from five distinct multi-component sources, resulting in a final catalog of 93 entries. The positions, peak brightness, total flux densities, and primary beam corrections from these radio sources are reported in Table \ref{tab:VLA sources}. 
\begin{table*}[]
\centering
\begin{tabular}{ccc}
\hline
\hline
Parameter & Value & Description                                                                                                      \\ \hline
\texttt{mean\_map}           & default        & \begin{tabular}[c]{@{}c@{}}Determines how the background mean map is computed\\  and how it is used further.\end{tabular} \\
\texttt{adaptive\_rms\_box}   & False          & \begin{tabular}[c]{@{}c@{}}Defines an adaptive box to calculate the rms\\  and mean maps.\end{tabular}                    \\
\texttt{atrous\_do}          & False          & If True, wavelet decomposition will be performed.                                                                         \\
\texttt{rms\_box} &
  None &
  \begin{tabular}[c]{@{}c@{}}Calculates the rms and mean over the entire image.\\  Defines the number of pixels by which this box is moved\\  for the next measurement.\end{tabular} \\
\texttt{thresh\_isl}         & 3            & Determines the region to which fitting is done.                                                                            \\
\texttt{thresh\_pix}         & 5              & \begin{tabular}[c]{@{}c@{}}Sets the source detection threshold in number of sigma\\  above the mean.\end{tabular}         \\
\texttt{flag\_maxsize\_fwhm} &
  0.2-0.3 &
  \begin{tabular}[c]{@{}c@{}}Any fitted Gaussian whose contour of \texttt{flag\_maxsize\_fwhm}\\  times the FWHM falls outside the island is flagged.\end{tabular} \\
\texttt{group\_tol} &
  1-5 &
  \begin{tabular}[c]{@{}c@{}}Sets the tolerance for grouping of Gaussians\\  into sources: larger values will result in larger sources.\end{tabular} \\ \hline
\end{tabular}
\caption{Parameters of PyBDSF with their corresponding values used to extract the sources. }
\label{tab:parameters_PyBDSF}
\end{table*}

Some emission from the extended radio sources composed of multiple components (such as FR radio galaxies) was not properly modeled. To properly fit their extended emission and derive their total flux densities, the parameters \texttt{flag\_maxsize\_fwhm} and \texttt{group\_tol} were increased.

To evaluate the reliability of the detected radio sources, we derived the expected fraction of spurious sources in our catalog. We produced an inverted map by multiplying the 6\,GHz continuum map by -1. Following this, we executed \texttt{PyBDSF}  employing the same parameters used to generate the radio source catalog. This process resulted in the detection of 15 spurious sources with peak $\rm SNR \simeq 5-6$, leading to a total fraction of spurious sources of 16\% and a purity $p > 0.83$  \citep[e.g.,][]{gonzalez2020alma, gomez2022goods,  fujimoto2023dualz}, defined as
\begin{equation}
    p = \frac{N_{\rm{pos}} - N_{\rm{neg}}}{N_{\rm{pos}}},
\end{equation}
where $N_{\rm{pos}}$ and $N_{\rm{neg}}$ represent the number of genuine and spurious sources at a given SNR, respectively.

\subsection{Radio Size Estimates}

\texttt{PyBDSF} provides information on the deconvolved FWHM of the major/minor axis of the radio sources and their uncertainty. Here, we elaborate on how the deconvolved FWHM, i.e., the intrinsic extent of the radio sources,  and associated error were derived.  In the case of a circular beam, the deconvolved FWHM is given by
\begin{equation}
    \theta = \left(\phi^2 - \theta_{1/2}^2\right)^{1/2} ,
\end{equation}
where $\phi$ is the FWHM of the fitted major or minor axis of the sources and $\theta_{1/2}$ is the FWHM of the synthesized beam, in our case  $\theta_{1/2} = 0\farcs82$.

The uncertainties in the deconvolved FWHM were computed as in \citet{murphy2017goods} using 
\begin{equation}
    \left(\frac{\sigma_{\theta}}{\sigma_{\phi}}\right) = \left[1- \left(\frac{\theta_{1/2}}{\phi}\right)^2\right]^{-1/2},
\end{equation}
where $\sigma_{\phi}$ is the uncertainty of the FWHM before deconvolution. 

To consider a source as confidently resolved along the major axis, we followed the criterion $\phi_M - \theta_{1/2} < 2 \sigma_{\phi_M}$ as in  \citet{murphy2017goods, heywood2021vla},  where $\phi_M$ and $\sigma_{\phi_M}$ are the major axis FWHM of the source before deconvolution and its associated error (provided by \texttt{PyBDSF}). Around $39\%$ of the radio sources in our sample are reliably resolved, i.e.,  36 out of the 93 sources in the catalog; see Figure \ref{fig:selection_function}.  
Consequently, our data set is mainly composed of upper limits to the galaxy sizes that remain unresolved in our 6\,GHz map. When a source reports a FWHM = 0, the upper limit is taken as the 1$\sigma$ error of the fitted FWHM that is $\approx0.1$ arcsec. To estimate the median properties, we employed survival analysis using the Kaplan–Meier \citep[KM;][]{kaplan1958nonparametric} estimator as implemented in the Python package \citep[\texttt{lifelines};][]{Davidson-Pilon2019}. This approach incorporates the censored observations (i.e., the upper limits) to reconstruct the true underlying distribution in a maximum‑likelihood‑style framework \citep{feigelson1985statistical}.

\subsection{Selection Function}
To infer the selection function imposed by the radio map properties and our source extraction, we derived the maximum detectable angular size as a function of total flux density. This is done by using the relation \citep[see Appendix C of][]{murphy2017goods}
\begin{equation}\label{eq: speak/sint ratio}
    \frac{S_{\rm{peak}}}{S_{\rm{int}}} = 2\rho^2[1-\sqrt{\pi}\rm{z} \exp(\rho^2)\rm{erfc}(\rho)],
\end{equation}
where $\rho\approx 0.50398(\theta_{1/2}/R_{\rm{eff}})$, with $\theta_{1/2}$ the FWHM of circular beam, and $R_{\rm{eff}}$ the exponential effective radii approximated as $R_{\textup{eff}} = \theta_M / 2.430$ \citep{murphy2017goods}, where $\theta_M$ is the deconvolved FWHM provided by \texttt{PyBDSF}.
Then, considering that $S_{\rm{peak}} = 5\,\mu$Jy i.e., our detection threshold, Equation \ref{eq: speak/sint ratio} was solved for $R_{\rm{eff}}$ using the Newton-Raphson method with the \texttt{scipy.optimize} library in Python. 

\begin{figure}
    \centering
    \includegraphics[scale=0.41]{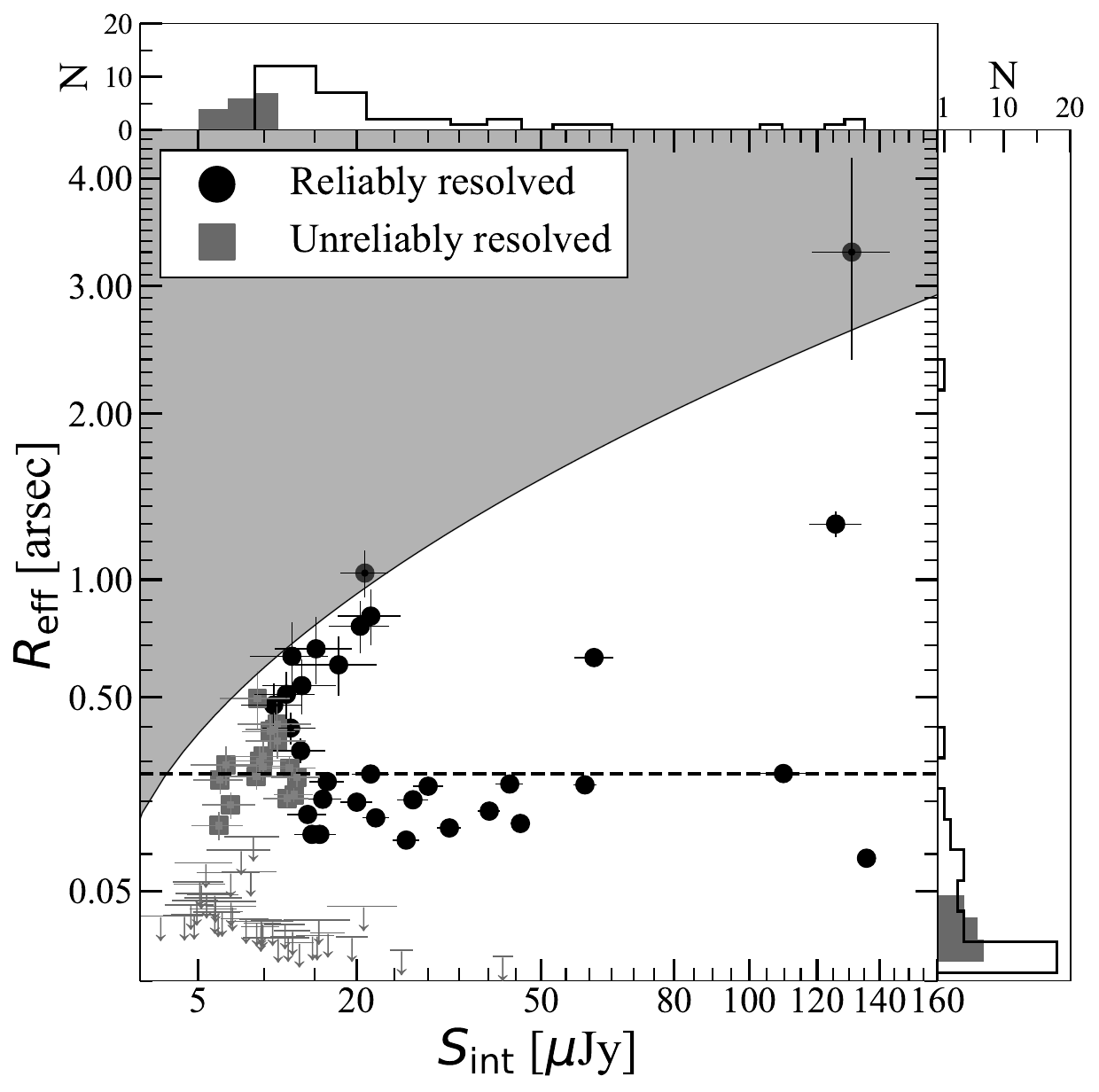}
    \caption{Maximum detectable angular size as a function of the integrated/total flux density of sources in the 6\,GHz image of A\,2744. Reliably resolved sources are shown in black circles, whereas unreliably resolved sources are presented in gray squares. The gray shaded region shows the selection function area. The horizontal dashed line is located at the median effective radius of 0\farcs267. The vertical arrows are upper limits to the 6\,GHz effective radius of sources that are not reliably resolved. In the top panel, counts of reliably resolved sources (black) and unreliably resolved sources (gray) are shown as a function of $S_{\rm{int}}$. On the right panel, counts of reliably resolved sources (black) and unreliably resolved sources (gray) are shown as a function of detectable angular size.}
    \label{fig:selection_function}
\end{figure}

The selection function (see Figure \ref{fig:selection_function}) imposed by the resolution and sensitivity of our 6\,GHz image of A\,2744 shows that we probe sources as faint as $\sim 6.5\,\mu$Jy with a minimum and maximum FWHM of $\sim 0\farcs1$ and $\sim 3\farcs5$, respectively. Close to the detection limit, we tend to detect compact sources that, due to their faint nature, are unreliably resolved. On the contrary, the reliably resolved sources are extended and have higher flux densities. We used a survival analysis with the Kaplan–Meier estimator that includes upper limits to the radio size of unresolved sources (FWHM = 0) to obtain a median effective radius $0\farcs267$ and 25th/75th percentiles of $0\farcs198/0\farcs495$. The median flux density is $14.7\,\mu$Jy\,beam$^{-1}$ and 25th/75th percentiles of 11.2/26.7\, $\mu$Jy\,beam$^{-1}$ (see histograms in Figure \ref{fig:selection_function}). Considering the radio sources with redshift values (see Section \ref{section: Counterpart asociation}), we derived the physical size from the angular sizes extracted by \texttt{PyBDSF} ---those values are reported in Table \ref{tab:VLA sources}. 

\section{\textit{JWST + HST} counterpart association}\label{section: Counterpart asociation}
We cross-matched our VLA 6\,GHz catalog with the ``UNCOVER Photometric Catalog'' presented in \cite{2024ApJS..270....7W}, which includes all the available JWST/NIRCam imaging and HST ACS/WFC3 deep observations of A\,2744, see Table \ref{tab:JWST_HST_programs}.

Using a search radius of $0\farcs5$, resembling the angular resolution of the 6\,GHz image, within the area where the VLA footprint overlaps with the JWST coverage, only 48 VLA radio sources are present. Among these, 46 (approximately 96\%) have candidate counterparts identified in the JWST mosaics. In cases where multiple sources fell within the search radius  (2 instances), the nearest one was adopted as the counterpart. The largest separation among these associations was $0\farcs43$. Each source was visually inspected to confirm the correct counterpart identification. From the 2 remaining sources without counterparts, the source N.54 has a peak brightness of $5.07\,\mu$Jy corresponding to an $\rm SNR \sim 5$ and the source N.91 reports a peak brightness $\sim 5.35\,\mu$Jy corresponding to an SNR$\sim 5$. Both sources are not detected in the VLA L- and S-band maps from \citet{pearce2017vla}. According to the analysis presented in Section \ref{section:Source extraction}, the low significance measured in both cases initially suggests that the detections could be spurious, albeit a scenario in which this source is a dust-enshrouded galaxy can not be ruled out.  Even if are spurious, the fraction of spurious sources in the subset of 48 VLA sources within the JWST coverage is lower ($\approx 4\%$), suggesting that the reliability of our catalog is higher than expected from the negative source count analysis (Section \ref{section:Source extraction}).  

\begin{figure*}
    \centering
    \includegraphics[width=1\linewidth]{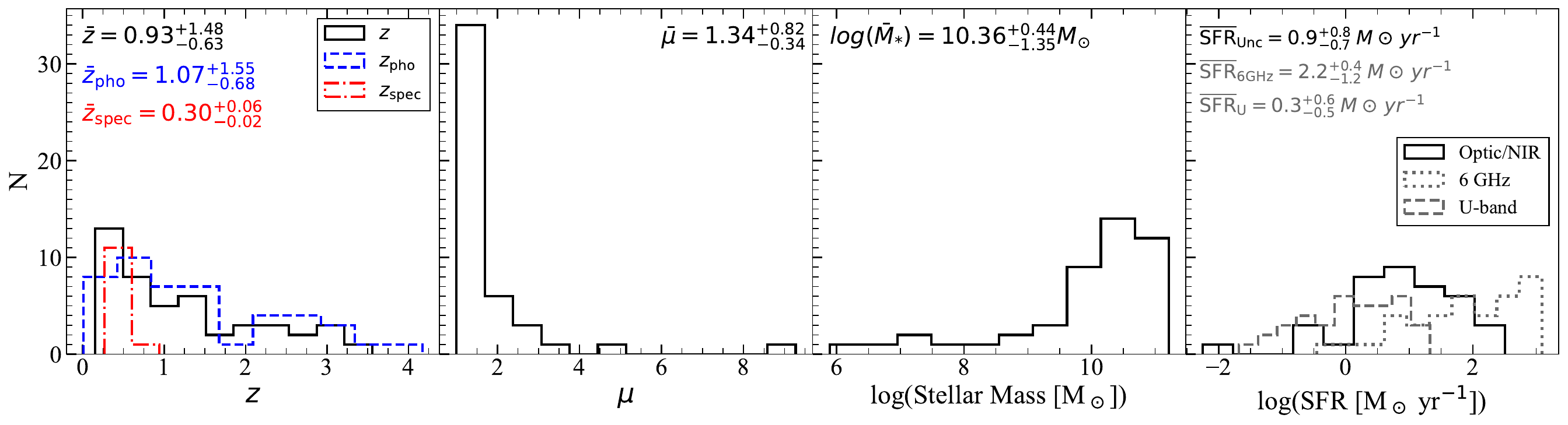}
    \caption{Histograms of redshift ($z$), magnification ($\mu$), stellar mass ($M_\star$), and SFR values reported by the UNCOVER survey \citep{2024ApJS..270....7W} for the 46 VLA radio sources with a JWST/HST counterpart. The upper and lower values are the corresponding 16th and 84th percentiles. In the first panel, the black solid line corresponds to the redshift distribution considering $z_{\rm{phot}}$ and $z_{\rm{spec}}$. The blue dashed line corresponds to the photometric redshift distribution, and the red dashed line corresponds to the spectroscopic redshift distribution. In the SFR histogram, the optical/NIR SFRs distribution reported by \cite{wang2023uncover}, are shown in a black solid line, while the distribution of the radio-based SFRs derived in this work is shown as a grey solid line, and the $u$-band SFR is given by the dashed grey line. AGN candidates are excluded from the histogram of SFR estimates (see Section\,\ref{subsec: AGN fraction}).}
    \label{fig:histogram_of_z_mu_ms_sfr}
\end{figure*}

The counterpart association process yielded values for redshift ($z$), magnification ($\mu$), stellar mass ($M_\star$), and SFR for the 46 VLA radio sources with a JWST/HST counterpart, see Table \ref{tab:counterparts _data}. Those values were derived by \cite{wang2023uncover} via spectral energy distribution (SED) fitting using the \texttt{Prospector} Bayesian inference framework \citep{johnson2021stellar}, with two notable modifications. First, observationally motivated priors on stellar mass, metallicity, and star formation history (SFH) from Prospector-$\beta$ were optimized to improve photometric redshift accuracy \citep{wang2023uncover}. Second, the magnification–redshift relationship was solved within \texttt{Prospector} using mass-dependent priors. The fits were performed using the simple stellar populations (SSPs), which come from FSPS \citep{conroy2010propagation}, with MIST isochrones \citep{choi2016mesa, dotter2016mesa} and
MILES stellar library \citep{bean2022casa}.
The composite stellar populations (CSPs) were modeled with Prospector-$\beta$ \citep{wang2023uncover}, and dust emission was included in all fits \citep{draine2007infrared}. The attenuation of the intergalactic medium (IGM) was assumed to follow \cite{madau1995radiative}. The corresponding $z$, $\mu$, $M_\star$, and SFR distributions are presented in Figure \ref{fig:histogram_of_z_mu_ms_sfr} and further discussed in Section \ref{subsec:magnificarion, redshift and stellar mass}. 

\begin{figure}
    \centering
    \includegraphics[width=1.1\linewidth]{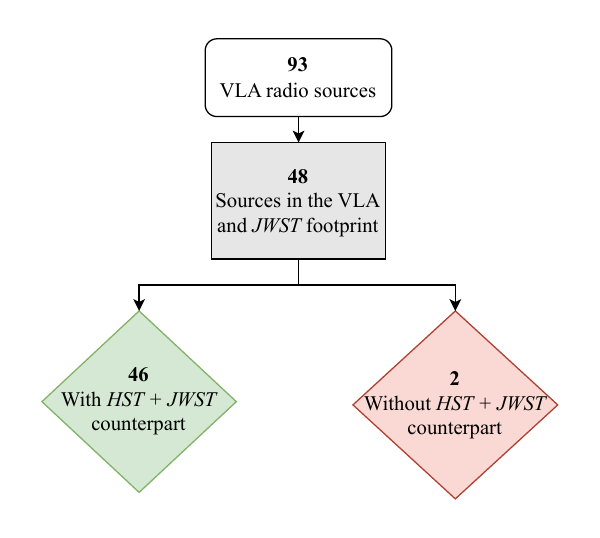}
    \caption{A flowchart that illustrates our matching procedure to identify VLA radio sources with and without counterparts in \textit{JWST} and \textit{HST} imaging. The green color is related to the VLA sources with counterparts and the red color is related to the VLA sources without counterpart. In summary, 46 sources have counterparts, 2 are likely spurious radio sources, and 45 lie outside the JWST coverage.}
    \label{fig:sche_counterpart_association}
\end{figure}

The summary of the counterpart association process is presented in a flowchart view in Figure \ref{fig:sche_counterpart_association}.
We produced RGB images for the 46 radio sources with JWST + HST counterparts, using the JWST NIRCam filters R: F444W, G: F277W, B: F150W, and overlaid contours of VLA emission; see the captions in Figure \ref{fig:RGB_Images_2} for more information.

\section{Results}\label{sec: Results_section}
\subsection{Magnification, Redshift, Stellar Mass, and Size of Galaxies}\label{subsec:magnificarion, redshift and stellar mass}
\begin{figure}
    \centering
    \includegraphics[width=1\linewidth]{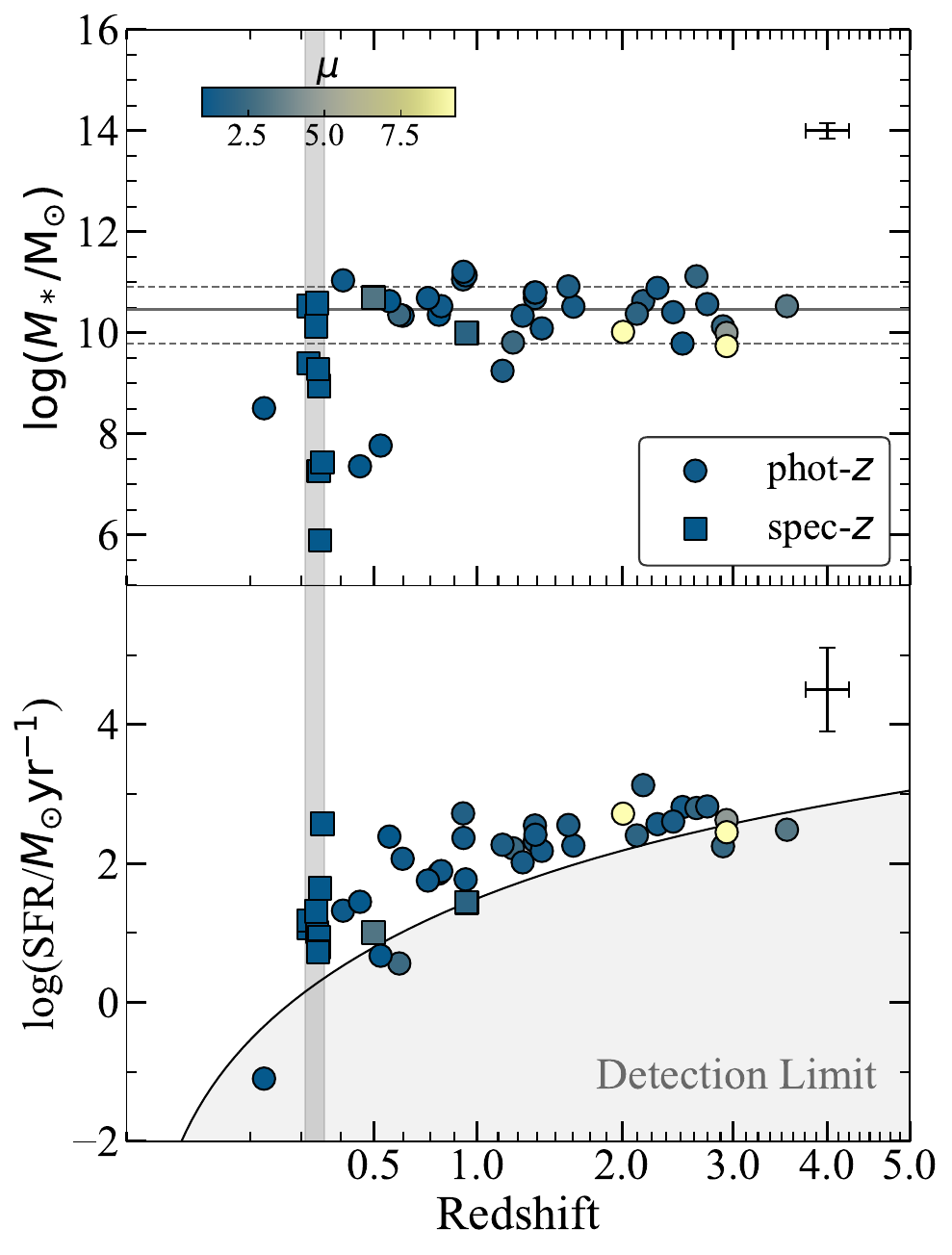}
    \caption{Top panel: Stellar mass as a function of redshift for the 46 VLA galaxies with HST+JWST counterparts. The solid and dashed lines correspond to the 50th, 84th, and 16th percentiles of the stellar mass distribution. Representative error bars of $\pm0.25$\,dex in stellar mass and $\pm 0.3$ in redshift are shown. Bottom panel: Star formation rate as a function of redshift. The grey shaded region shows our detection limit without magnification, defined by the SFR of a galaxy detected at the $5\sigma$ level ($\sim 5.45\,\mu$Jy beam$^{-1}$). Error bars represent an uncertainty of 0.75 dex in SFR and $\pm 0.3$ in redshift. In both panels, squares represent sources with spectroscopic redshifts and circles represent galaxies with photometric redshifts. The symbols are colored according to the source magnification. The vertical shaded region shows the cluster members. These plots illustrate the advantage of using gravitational lensing to detect less massive/active galaxies at higher redshifts ($z \gtrsim 1$) that could not have been detected without magnification factors larger than 1.}
    \label{fig:M_star vs redshift and SFR vs redshift}
\end{figure}

For the 46 sources with JWST + HST counterparts, the median values of the physical parameters are reported in Table \ref{tab:result_values}. Only a small fraction ($\approx 24\%$) are moderately lensed, with magnification factors greater than 2 (Figure \ref{fig:M_star vs redshift and SFR vs redshift}). The source with the highest magnification factor ($\mu = 9.88$) in our sample is the N.62 with a peak brightness of $8.4 \pm 1.2\,\mu\,\rm Jy\,beam^{-1}$ and is located at $z_{\rm{phot}} \approx 2.94^{+0.05}_{-0.04}$.  The radio-based SFR of this source is $180^{+50}_{-40}\,\rm{M}_{\odot}\rm{yr}^{-1}$ with a stellar mass of $\approx 5.5_{-1.0}^{+1.2}\times10^{9}\,\rm{M}_{\odot}$, corresponding to a starburst galaxy that lies above the main sequence of SFGs from \cite{leslie2020vla}.
From the 46 sources, 12 (26\%) of them have spectroscopic redshifts. The source with the highest redshift is N.47 with ($z_{\rm{phot}} \approx 3.55^{+0.16}_{-0.11}$), peak brightness of $6.62 \pm 2.45\,\mu$Jy$\rm \,beam^{-1}$, and  SFR $\approx 304^{+140}_{-100}\,\rm{M}_{\odot}\rm{yr}^{-1}$. 
\begin{table}[]
\centering
\begin{tabular}{cccc}
\hline
\hline
Property               & Symbol & Value & Units \\ \hline
Magnification factor   & $\bar{\mu}$     & $1.34^{+0.82}_{-0.34}$     & -     \\
Redshift               & $\bar{z}$      & $0.93^{+1.48}_{-0.63}$     & -     \\
Spectroscopic redshift & $\bar{z}_{\rm{spec}}$      & $0.30^{+0.06}_{-0.02}$     & -     \\
Photometric redshift   & $\bar{z}_{\rm{phot}}$      & $1.07^{+1.55}_{-0.68}$     & -     \\
Stellar mass           & $\log(M_\star/\rm{M}_{\odot})$  & $10.36^{+0.44}_{-1.35}$     & $\rm{M}_{\odot}$ \\
Physical size              & $\bar{D}_{\rm{phys}}$   & $1.5^{+1.8}_{-0.4}$     & kpc   \\ 
Star formation rate             & $\overline{\rm{SFR}}_{\rm{6\,GHz}}$   & $2.2^{+0.4}_{-1.2}$     & $\rm{M}_{\odot}\rm{yr}^{-1}$   \\ \hline
\end{tabular}
\caption{Physical properties of the 46 radio sources with  \textit{JWST + HST} counterparts. The median redshift ($\bar{z}$) refers to the combined spectroscopic and photometric values. The median star formation rate ($\overline{\rm{SFR}}_{\rm{6,GHz}}$) is estimated from our 6\,GHz fluxes.}
\label{tab:result_values}
\end{table}
The absence of sources in the sampled mass regime below $\log(M_\star/\rm{M}_{\odot}) \lesssim 9.5$ (see\,Figure \ref{fig:M_star vs redshift and SFR vs redshift}) at higher redshifts ($z\gtrsim 1$) is a result of our radio detection limit that preferentially selects massive, bright systems. 
The sizes of the galaxies at 6\,GHz in the three Hubble Frontier Fields (MACS J0416.1-2403, MACS J0717.5+3745, and MACS J1149.5+2223) have been previously explored by \cite{jimenez2021vla}. They reported a median effective radius of $\bar{R}_{\rm{eff}} = 1.1^{+0.7}_{-0.3}$\,kpc for the 31 radio-detected galaxies  with $\log(M_{*}/\rm{M}_{\odot}) \approx 10.4$ distributed over $0.3 \lesssim z \lesssim 3$. Our values are consistent with those reported by \cite{jimenez2021vla}, which is expected given that our sample is very similar in terms of redshift and stellar masses, and that we used comparable resolution and sensitivity.
\subsection{Radio Spectral Index}
To derive robust star formation rates, we first calculated the spectral indices ($\alpha$) using S-band, L-band, and C-band flux densities from \citet{pearce2017vla} and our work. To ensure consistent photometry across different frequencies, the C-band and S-band maps were convolved to the lowest angular resolution (L-band map). Sources were subsequently extracted to obtain matched-aperture fluxes. We identified 24 sources with clear counterparts across all bands. The spectral indices were determined by performing a linear least-squares fit in log-log space ($S_\nu \propto \nu^\alpha$). We obtained a median spectral index of $\alpha \approx -0.8$, which is consistent with the typical values for star-forming galaxies ($\alpha \approx -0.7$) reported in the literature \citep{condon1992radio, smolvcic2017vla}. These results are summarized in Table~\ref{tab:counterparts _data}.

\subsection{Identifying AGN Candidates}\label{subsec: AGN fraction}
We have to consider that AGN might be present in our 6\,GHz radio source catalog of A\,2744. Based on the numerical simulations and models of \cite{mancuso2017galaxy} and \cite{bonaldi2019tiered}, the AGN fraction in radio surveys at $\approx$\,6\,GHz detecting sources with total flux densities $\gtrsim 5\,\rm \mu Jy$ is $\approx14$\%. Thus, our radio source catalog is expected to be dominated by SFGs. 

To confirm such predictions and identify AGN candidates in our 6\,GHz radio source catalog of A\,2744, we review the \citet{labbe2024uncover} work, which searches for AGN candidates in A\,2744 but only has results in a redshift range of $3<z<7$.
We then implemented two AGN diagnostics specifically for the sub-sample of 46 VLA sources that have confirmed counterparts and available redshifts ($z_{\rm{spec}}$ and/or $z_{\rm{phot}}$) reported in Table \ref{tab:counterparts _data}. A source in the 6\,GHz catalog was deemed as an AGN if it has an X-ray luminosity $L_{\rm{X}} > 10^{42}$\,erg s$^{-1}$; and/or exhibits an excess of radio emission as expected from the IR-radio correlation of SFGs. While X-ray detections could potentially identify AGN in the full radio sample, we restrict this analysis to these 46 sources where the X-ray luminosity and the infrared-radio correlation can be robustly constrained.

\begin{figure}
    \centering
    \includegraphics[width=1\linewidth]{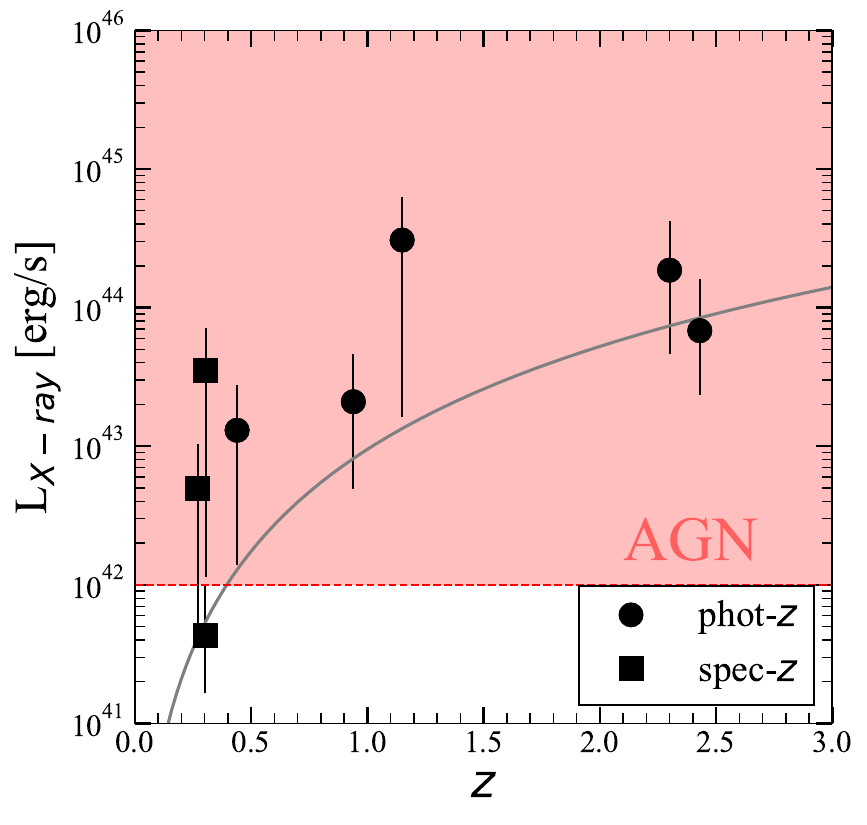}
    \caption{Left: X-ray luminosities of the VLA sources with X-ray counterparts. The red-shaded region shows the parameter space ($L_{\rm{x}} > 10^{42}$ erg s$^{-1}$) dominated by AGN. The black solid line corresponds to the detection limit of the broad band (0.5-7 keV). Circles and squares denote VLA sources with photometric and spectroscopic redshifts, respectively.} 
    \label{fig:AGN_candidates}
\end{figure}

We used the ``The Chandra Source Catalog (CSC)'' survey \citep{evans2018chandra} to find X-ray fluxes in the broadband (0.5-7 keV) of our VLA radio sources. Using a cross-match radius of $1\farcs0$, we found 8 X-ray counterparts, of which 7 meet the first criterion ($L_{\rm{X}} > 10^{42}$\,erg s$^{-1}$) and are deemed as AGN candidates (see Figure \ref{fig:AGN_candidates}). The source N.40, which shows the lowest X-ray luminosity, could be an AGN when considering the uncertainties and absorption. However, in this work, we will still consider it as a star-forming galaxy.

For the second criterion, we searched for IR counterparts using the HST Frontier Fields Herschel catalog presented by \citet{rawle2016complete}.  A search radius of $3\farcs0$ led to the identification of 6 IR counterparts whose  total IR luminosity ($L_{\rm{IR}}$), integrated over the rest-frame wavelength range $\lambda = 8-1000 \mu$m, spans in a range of $ 9.3\lesssim \log(L_{\rm{IR}}/L_{\odot})\lesssim 12.6$ and are reported by  \citet{rawle2016complete}. We  then inferred the expected radio luminosity from the infrared-radio correlation parameterized as \cite{helou1985thermal}
\begin{equation}
    q_{\rm{IR}} = \log \left(\frac{L_{\rm{IR}}}{3.75\times 10^{12}\,\rm{Hz}}\right) - \log \left(\frac{L_{\rm{1.4\,GHz}}}{\rm{{W Hz}^{-1}}}\right),
\end{equation}
where $L_{\rm{IR}}$ is the integrated IR luminosity over the rest-frame wavelength range $\lambda = 8-100 \mu$m reported by  \citet{rawle2016complete}. While, $L_{\rm{1.4\,GHz}}$ is the 1.4\,GHz monochromatic radio luminosity.

To identify the radio-excess sources, we use the IRRC model for star-forming galaxies reported by \cite{delhaize2017vla},
\begin{equation}
    q_{\rm{IR}}(z) = (2.88 \pm 0.03)(1 + z)^{-0.19 \pm 0.01}
\end{equation}
Following  \cite{radcliffe2021nowhere}, the VLA sources with $q_{\rm{IR}}$ below the mentioned $3\sigma$ error bounds are classified as radio excess sources (i.e., AGN candidates). From this criterion, 2 AGN candidates were found, one of them (N.26) also shows an X-ray luminosity excess fulfilling the first criterion.  

Beyond the primary sub-sample of 46 sources, a visual inspection identified that source N.1 is a clear radio galaxy exhibiting a prominent radio lobe, confirming its AGN nature. Including this morphologically identified source, we find a total of 9 AGN candidates. Consequently, the AGN fraction for the sample of sources with either available redshifts or clear morphological identification is  {9/46 ($\sim$ 20\%)}, which is comparable with the predicted AGN fraction of 14\% in $\rm \mu Jy$-level radio surveys at intermediate frequencies \citep[$\approx6$\,GHz; ][]{mancuso2017galaxy, bonaldi2019tiered}. 
\subsection{Deriving  Unobscured  and Obscured SFRs }\label{sub_sec: deriving SFR}
After removing the AGN candidates from our sample, we derived the SFR from the  6\,GHz radio emission that is tracing dust-obscured star formation. We used the SFR calibrations from \cite{murphy2017goods}, which were normalized  to a Chabrier initial mass function (IMF; as used by the UNCOVER team),
\begin{equation}
    \left(\frac{\rm{SFR}_{6\,\rm{GHz}}}{\rm{M}_{\odot} \rm{yr}^{-1}}\right) = 4.87 \times 10^{-22}\left(\frac{L_{1.4\,\rm{GHz}}}{\rm{W}\,\rm{Hz}^{-1}}\right).
\end{equation}
$L_{\rm{1.4\,GHz}}$ is given by
\begin{equation}
    L_{\rm{1.4\,GHz}} = \frac{4\pi [D_L(z)]^2}{(1+z)^{1-\alpha}}\left(\frac{1.4}{6}\right)^{-\alpha} S_{\rm{6\,GHz}},
\end{equation}
where $S_{\rm{6 GHz}}$ is the 6\,GHz flux density in W\,m$^{-2}$\,Hz$^{-1}$, $D_L$ is the luminosity distance in m, and $\alpha$ is the spectral index. For sources detected in multiple bands (L$-$ and/or S$-$band), we adopted the empirically derived spectral indices described above. Conversely, for sources with detections only in the C-band, we assumed a canonical spectral index of $\alpha = -0.7$, with an intrinsic dispersion of $\sim 0.1$ \citep[e.g.,][]{condon1992radio, smolvcic2017vla, klein2018radio}.

Typically, SFRs that are only slightly obscured by dust are estimated using observations in H$\alpha$ and UV, as these tracers are sensitive to the light emitted by young, massive stars. In this work, we follow the \cite{hopkins2003star} calibration to estimate the unobscured SFR using $u$-band flux densities from HST observations, as reported by the UNCOVER team \citep{labbe2024uncover}. For galaxies at $z \gtrsim 1$, this approach provides a direct measurement of the rest-frame ultraviolet emission, effectively probing the UV continuum.

The unobscured SFR of galaxies was estimated using the rest-frame $u$-band flux densities reported from the UNCOVER survey by \cite{wang2023uncover}. We can use the relation \citep{hopkins2003star}:
\begin{equation}
        \left(\frac{{\rm SFR}_{u}}{\rm{M}_{\odot}\rm{yr}^{-1}}\right) = \left(\frac{L_{u}}{1.81 \times 10^{21}\, \rm{W}\,\rm{Hz}^{-1}}\right)^{1.186},
\end{equation}
 where $L_{u}$ is the $u$-band luminosity density. This calibration was obtained using a sample of $u$-band luminosities derived from the SDSS K-corrected absolute $u$-band magnitudes \citep{blanton2003estimating}, also an obscuration correction based on the Balmer decrement and the extinction curve from \cite{calzetti2001dust} has also been employed. Using this sample, the ordinary least-squares bisector method of linear regression \citep{isobe1990linear} was applied
to $\log(L_U)$ and $\log(\rm{SFR_{H\alpha}})$.

\subsubsection{Comparing Unobscured and Obscured SFRs}\label{subsection: comparison_between_SFR_stimates}

We compared the three SFR estimates available for the 46 sources with HST, JWST, and VLA counterparts and redshift information: the SFR derived via SED fitting using optical/near-infrared data reported by the UNCOVER team \citep{wang2023uncover}, hereafter $\rm SFR_{\rm UNCOVER}$, the SFR derived from the $u$-band luminosity (${\rm SFR}_{u}$), and the SFR inferred from the 6\,GHz flux density ($\rm SFR_{\rm 6\,GHz}$). 

\begin{figure}
    \centering
    \includegraphics[width=1\linewidth]{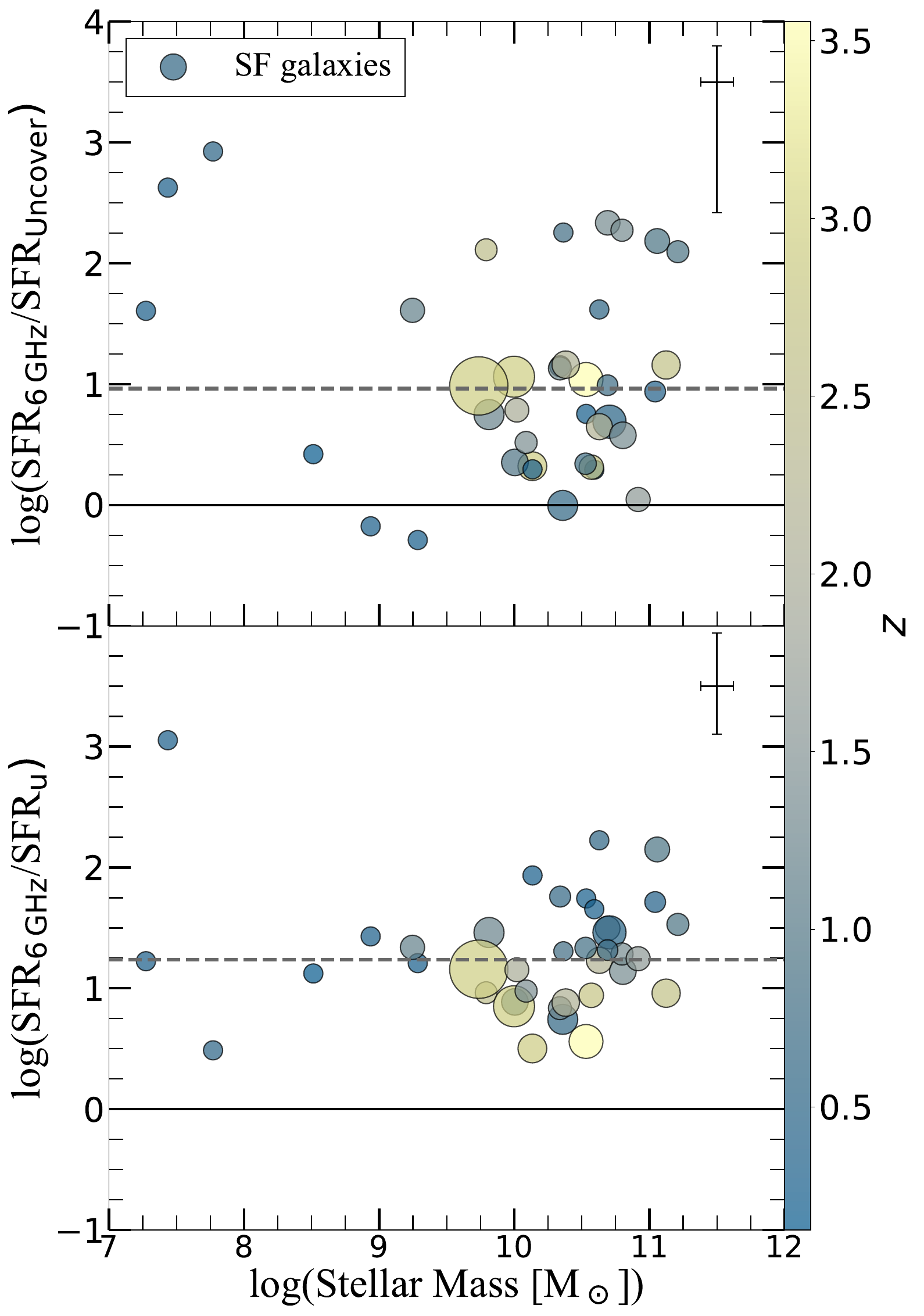}
    \caption{Top panel: Ratio of the $\rm SFR_{\rm 6\,GHz}$ to  $\rm SFR_{\rm Uncover}$  as a function of stellar mass. Representative error bars are shown: $\pm0.25$\, dex for stellar mass and $+0.29/-1.05$ dex for the SFR ratio. Bottom panel: Ratio of the $\rm SFR_{\rm 6\,GHz}$ to $\rm SFR_{u}$ as a function of stellar mass with representative error bars of $+0.41/-0.38$ dex and $\pm0.25$\, dex, respectively.
    In both panels, marker size scales with the magnification of each source, where larger markers indicate higher magnification. The color bar denotes the redshift. Solid horizontal lines indicate a 1:1 ratio between the radio-derived and the optical/near-IR or $u$-band star formation rates. The horizontal dashed lines represent the median ratios, showing that the 6\,GHz SFRs are typically higher by a factor of $\sim10$ compared to both the $u$-band and optical/near-IR estimates.}
    \label{fig:SFR_ratios_vs_stellar_mass}
\end{figure}

\begin{figure}
    \centering
    \includegraphics[width=1\linewidth]{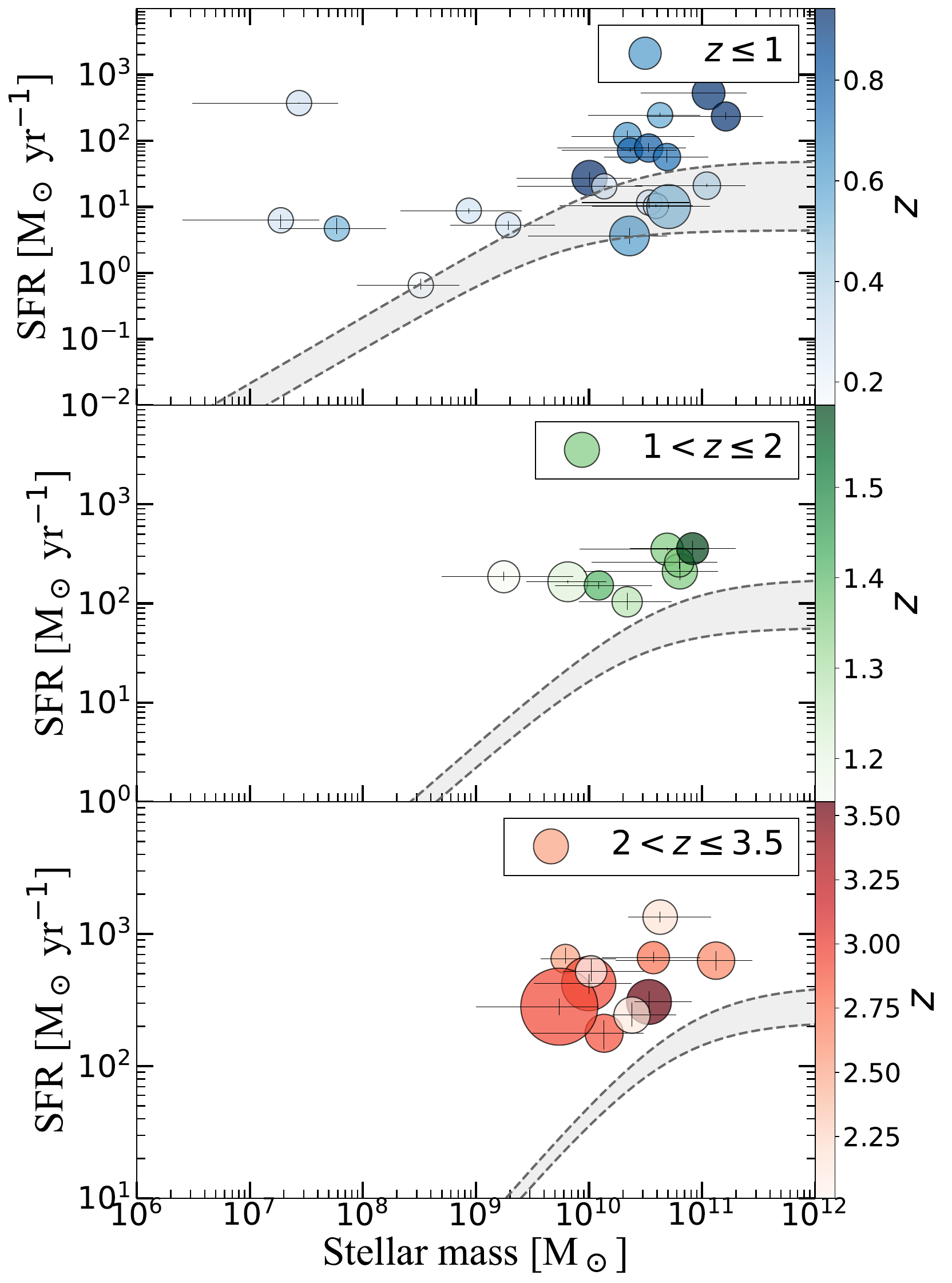}
    \caption{Sources in our sample in the SFR \textit{vs} stellar mass plane. The SFRs shown were derived from the 6\,GHz observations. From top to bottom, the first panel includes the sources with $0 \leq z \leq 1$, the second panel includes the sources with  $1 < z \leq 2$, and the third panel includes sources with $2 < z \leq 3.55 $. The gray shaded region represents the main sequence range following the \citet{leslie2020vla} model, evaluated at the mean redshift of each panel, defined by the midpoint between its minimum and maximum redshift. The magnification of each source is represented by the marker size; thus, larger makers indicate higher magnification. The AGN candidates are not plotted.}
    \label{fig:Main_sequence plane}
\end{figure}

When comparing the $\rm SFR_{\rm UNCOVER}$ with the $\rm SFR_{\rm 6\,GHz}$ derived in this work (see the left panel of Figure $\ref{fig:SFR_ratios_vs_stellar_mass}$), it is notable that the median $\rm SFR_{\rm 6\,GHz}$ is 10 times higher than the median $\rm SFR_{\rm UNCOVER}$. Such discrepancies can arise from a combination of effects, including the different timescales probed by each tracer \citep{2023A&A...675A.126A}, the assumed star formation histories \citep{2015MNRAS.447.3442L}, spatial mismatches and resolution effects, and uncertainties in energy-balance assumptions that may fail to recover the total SFR \citep{2011ApJ...738..106W}. In the specific comparison presented here, the stellar population templates adopted by \citet{wang2023uncover} to estimate SFRs are likely to be particularly uncertain in regimes of high stellar mass and low metallicity.
As observed in Figure \ref{fig:Main_sequence plane}, our sample is mainly composed of massive SFGs. 
More importantly, the discrepancy between $\rm SFR_{\rm UNCOVER}$ and $\rm SFR_{\rm 6\,GHz}$ suggests that the correction for dust extinction applied during the SED fitting is insufficient to account for the star formation activity that is heavily obscured in our massive SFGs.
 This highlights the importance of radio observations as star formation tracers since they are not affected by dust attenuation.
 
A similar discrepancy is observed when we compared the $\rm SFR_{\rm 6\,GHz}$ with those inferred from the rest-frame $u$-band luminosities reported in the UNCOVER catalog (see the bottom panel of Figure $\ref{fig:SFR_ratios_vs_stellar_mass}$). The median $\rm SFR_{u}$ is lower than the median $\rm SFR_{\rm 6\,GHz}$ by a factor of 50.
The discrepancy between the $u$-band and 6\,GHz SFRs can arise from several factors.  For example, the $u$-band luminosity can vary significantly during the stellar evolution, making it less reliable as an SFR tracer \citep{hopkins2003star}. Moreover, the average $u$-band obscuration correction \citep{calzetti2001dust} ranges from a factor  3 at SFRs of 1 $\rm{M}_{\odot}\rm{yr}^{-1}$ up to about a factor  10 at SFRs of 100 $\rm{M}_{\odot}\rm{yr}^{-1}$. This, again, stresses the need for star formation tracers unaffected by dust, like radio continuum emission. Without the 6\,GHz data, we would be underestimating the SFR of galaxies in our sample, typically by a factor of 10, because most of the star formation out to $z \sim 5$ remains heavily obscured \citep[e.g.,][]{bouwens2020alma,zavala2021evolution}.

Our $\rm SFR_{\rm 6\,GHz}$ values are not free of uncertainties. Albeit we consider the uncertainties related to the spectral index for SFGs (typically $0.7 \pm 0.1$) and the 5\% error floor in the $z_{\rm{phot}}$ imposed by \citet{2024ApJS..270....7W}, several systematic uncertainties are contributing to the dispersion of the SFR {\it vs} radio luminosity calibration. For instance, no systematic errors in the empirical IR-radio correlation are being considered, nor are the uncertainties introduced by adopting a given IMF. 

\subsection{Galaxies in the Star Formation {\it vs} Stellar Mass Plane}
We adopted the star-forming main sequence (SFMS) model from \citet{leslie2020vla}, using the parameterization for star forming galaxies provided in their Table 1. This model is particularly well-suited for our analysis as it is derived from radio-based SFRs for a substantial sample $\sim$ half-million galaxies in the COSMOS2015 catalog \citep{laigle2016cosmos2015}. Note that this model uses the Chabrier IMF as the UNCOVER catalog and the SFR estimates reported here. The 46 VLA radio sources with \textit{JWST/HST} counterparts are plotted in the logarithmic SFR-stellar mass diagram (see Figure \ref{fig:Main_sequence plane}). To compensate for the redshift dependence of the main sequence and for visualization purposes, we split our sample into three redshift bins: $0<z\leq1$, $1<z\leq2$, and $2 < z \leq 3.55 $, containing  20, 8, and 10 galaxies, respectively. 
As observed in Figure\,\ref{fig:Main_sequence plane}, SFGs preferentially lie at the massive end of the main sequence, while a minor fraction of low-mass ($\log(M_\star/M_\odot)\lesssim 9 $) starburst galaxies (i.e., above the main sequence) are also detected.
This is a consequence of our 6\,GHz detection limit, which imposes a minimum SFR (as a function of redshift) that can be detected in our image. This leads to the detection of both low- and high-mass starbursts, as well as massive galaxies on the main sequence that harbor high SFRs. Furthermore, as seen in the second and third panels, we do not detect galaxies with stellar masses below $10^{9}\,\mathrm{M}_{\odot}$. The absence of low mass galaxies in our sample can be largely attributed to the current detection limits of our 6\,GHz census. The 6 GHz sensitivity imposes a lower limit on SFR at any redshift, and low-mass galaxies are unlikely to achieve an SFR that high. However, beyond observational incompleteness, physical factors may also play a role. Low-mass galaxies often exhibit reduced radio emission, becoming ``radio-dim'', due to the inefficient retention of cosmic rays within their shallower gravitational wells \citep[e.g.,][]{2010ApJ...717....1L}.

\begin{figure*}
    \centering
    \includegraphics[width=0.85\textwidth]{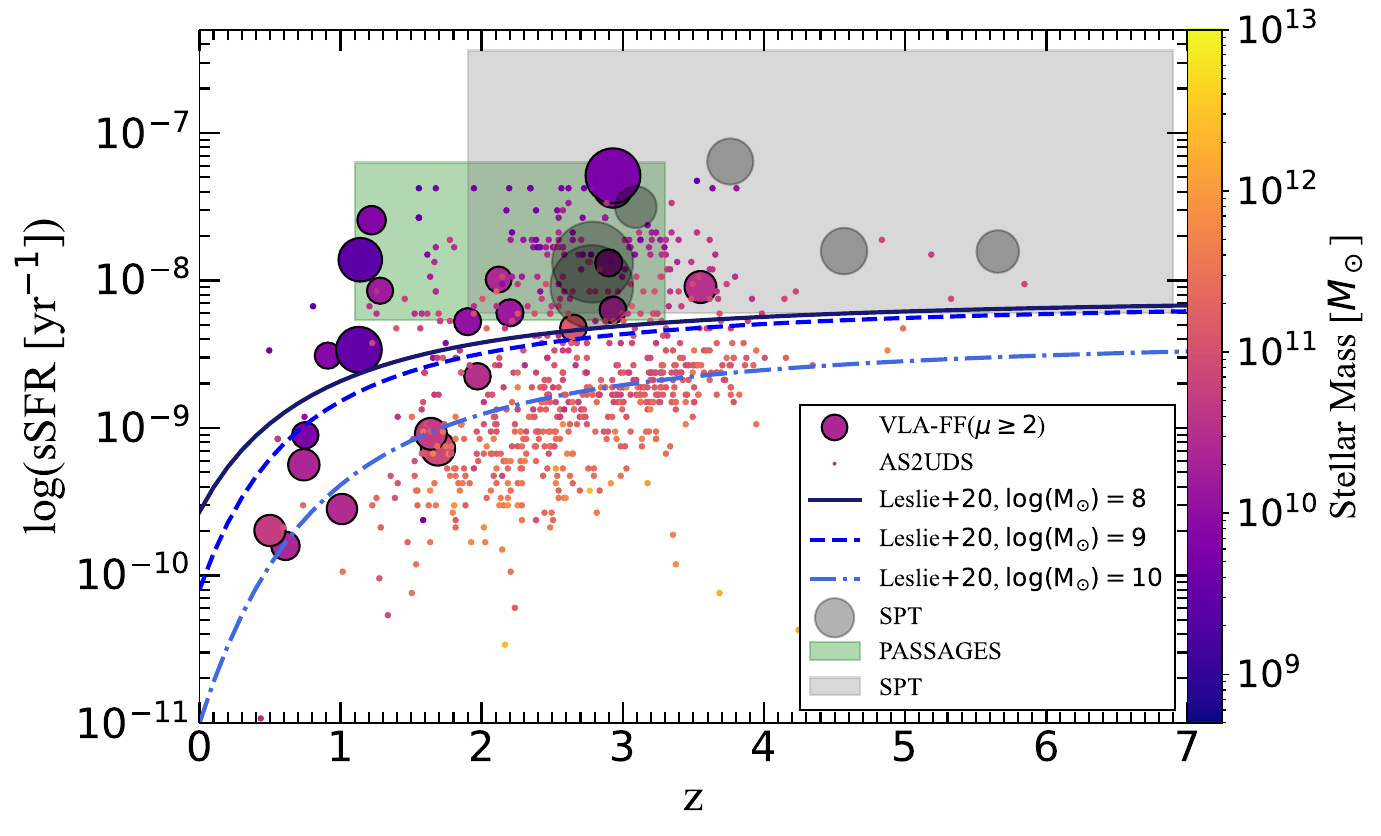}
    \caption{A sample of 22 moderately/strongly lensed ($\mu \geq 2$) sources from the VLA Frontier Fields program \citep[][and this work]{heywood2021vla} in the sSFR \textit{vs} redshift plane. The size is directly related with the magnification of each source. The gray hatched area shows the full sSFR–$z$ range of galaxies in the SPT survey \citep{reuter2020complete}, while the individual gray symbols mark the SPT galaxies reported by \citep{ma2015stellar}. The green hatched area indicates the sSFR–$z$ range covered by the PASSAGES survey \citep{kamieneski2024passages}. The little dots show the AS2UDS sample from \citep{dudzevivciute2020alma}. Solid lines trace the sSFR evolution from \citep{leslie2020vla} for galaxies with three representative stellar masses.}
    \label{fig:sfr_mass_magnified}
\end{figure*}

It is noteworthy that galaxies with higher magnification tend to have lower mass and SFR (see Figure \ref{fig:sfr_mass_magnified} and Table \ref{tab:counterparts _data}), emphasizing once again the importance of gravitational lensing for detecting fainter and more distant galaxies.

\subsection{Radio Properties of Little Red Dots}

A key discovery of the {\it JWST} was the abundant population of the so-called Little Red Dots \citep[LRDs; e.g.,][]{harikane2023jwst, kocevski2023hidden, labbe2023population, labbe2024uncover,barro2024extremely,kocevski2024rise}. These are  $z\gtrsim 4$ compact sources with red optical colors and frequently broad H$\alpha$ emission lines, suggesting the presence of type I AGN \citep[e.g.,][]{greene2024uncover,matthee2024little}. Alternatively, LRDs could be compact starbursts with  ionized outflows leading to emission line broadening \citep[e.g.,][]{wang2025rubies}. To gather a more complete view of the JWST-discovered LRDs, multi-frequency analysis have been implemented. Unexpectedly, it is found that the vast majority of LRDs are not detected in the deepest X-ray images \citep[][]{ananna2024x,yue2024stacking,maiolino2025jwst}, which is in conflict with expectations from broad line AGN scaling relations. In this context, radio observations are becoming relevant to trace the potential signatures of AGN processes. Several studies have looked for radio counterparts of LRDs identified in different cosmological fields \citep[e.g.,][]{mazzolari2024radio,perger2025deep}. Yet, only one LRD has been detected in radio \citep{gloudemans2025another}: PRIMER-COS 3866 at $z = 4.66$.

\begin{figure}
    \centering
    \includegraphics[width=1\linewidth]{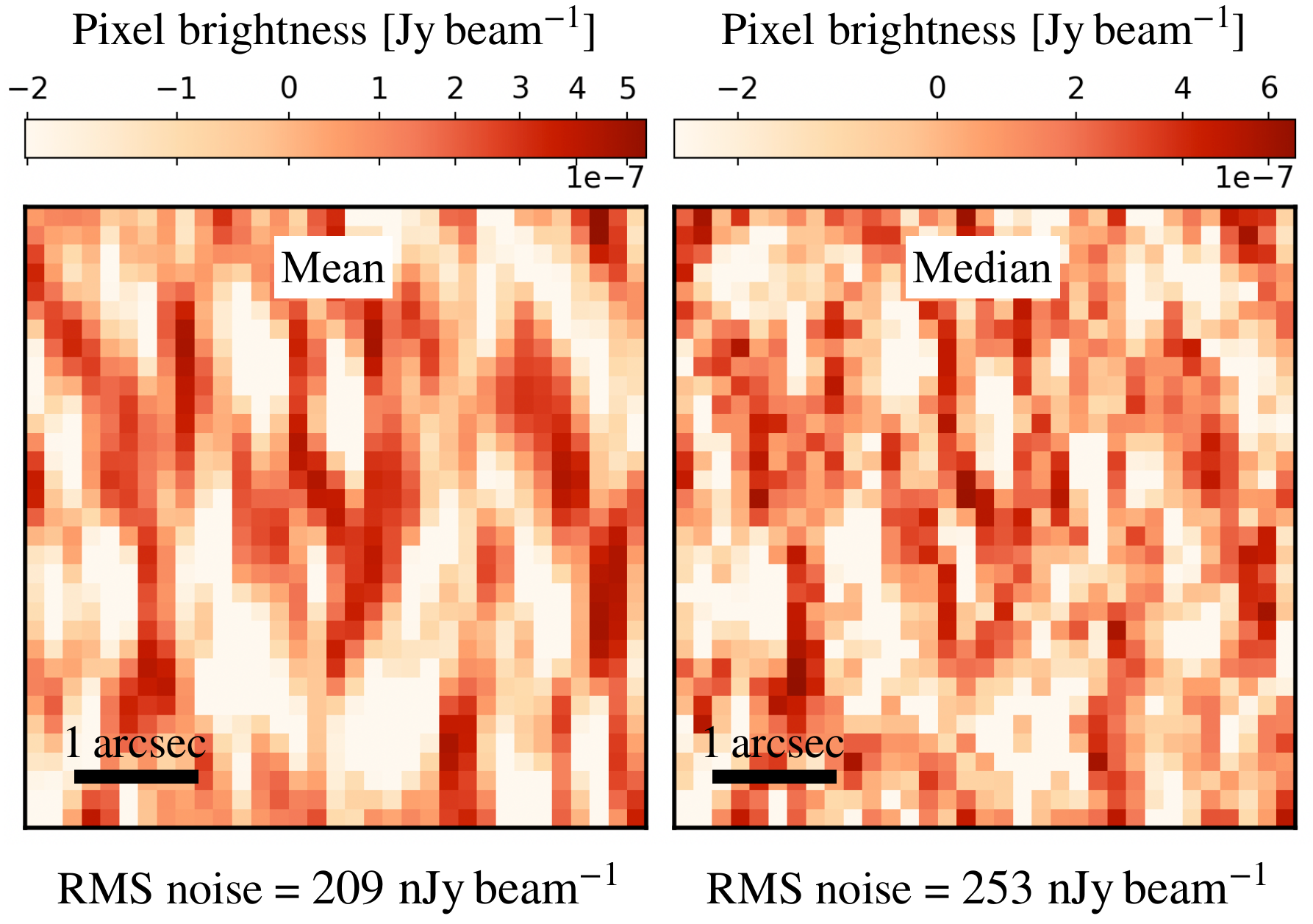}
    \caption{Mean and median stacked 6\,GHz images at the position of the 23 LRDs in A\,2744 reported by \citet{kocevski2024rise}. No signal has been detected down to an rms noise of $\approx 200\,\rm nJy\,beam^{-1}$.}
    \label{fig:stacked_lrd}
\end{figure}

Here, we crossmatched the catalog of LRDs reported by \citet{kocevski2024rise} with our 6\,GHz catalog of A\,2744. Despite the gravitational lens created by the cluster that increases the likelihood of detecting these puzzling high-$z$ sources, we found that none of the 23 LRDs in A\,2744 were detected in our map above peak $\rm SNR=5$. A visual inspection did not reveal any tentative detection at the position of the LRDs. After producing a stacked image of the 23 radio images, we found no significant detection. The resulting rms noise of the mean stacked image is $209\,\rm nJy\,beam^{-1}$ (or $253\,\rm nJy\,beam^{-1}$ if a median stack is adopted; see Figure\,\ref{fig:stacked_lrd}). Since we are stacking lensed galaxies, we corrected the rms noise values by lensing magnification by taking the mean (or median) $\mu$ value of the stacked galaxies, which is 1.95 (1.50). This leads to a 3$\sigma$ limit to the observed 6\,GHz  radio emission of $\approx 320\rm \,nJy\,beam^{-1}$ ($\approx 500\rm \, nJy\,beam^{-1}$). Note that at the median redshift of the sample  of $z\approx 6$, these observed flux density constraints correspond to rest-frame frequencies of $\approx 42\,\rm GHz$. Adopting a typical radio spectral index of $\alpha=0.7$, we derive a $3\sigma$ upper limit to the monochromatic luminosity, $\nu L_\nu$, at rest-frame 6\,GHz of $4.3\times 10^{39}\,\rm erg\,s^{-1}$ ($6.7\times 10^{39}\,\rm erg\,s^{-1}$). Such an upper limit is comparable with that reported for LRDs in the COSMOS and GOODS fields, whose radio luminosity at 1.3-5\,GHz in the rest frame is $\lesssim 2 \times 10^{39}\,\rm erg \,s^{-1}$ \citep{mazzolari2024radio,gloudemans2025another}. The expected radio luminosity of LRDs from constraints on their X-ray emission is $<  10^{37-39}  \,\rm erg \,s^{-1}$ \citep[see Section 5.1 of][]{gloudemans2025another}. 
 Since our upper limit to the radio luminosity  of LRDs is still close/above the expected value, deeper observations are needed to provide robust constraints on the origin of LRDs and their potential AGN.

\subsection{A Sample of Radio-selected, Moderately Lensed Galaxies in the Frontier Fields}
A key goal of the VLA Frontier Fields project is to leverage gravitational lensing to detect high-redshift and/or intrinsically faint galaxies. On this regard, here we present a compilation of 22 moderately/strongly lensed galaxies (with $\mu \geq 2$) in the VLA Frontier Field project. 13 of such galaxies are found in the MACSJ0416.1-2403, MACSJ0717.5+3745, and MACSJ1149.5+2223 fields \citep{heywood2021vla}, while 9 lie in the A\,2744 field (see  Table\,\ref{tab:fuentes_ssfr}.). The stellar mass and SFR range of this sample is $\log(M_\star/\rm{M}_{\odot})\approx 9.4 -11.2$ and $5-300\,\rm{M}_{\odot}$year$^{-1}$, respectively,  and span across a redshift and magnification range of $\approx 0.5-3.55$ and $\approx 2.06 - 9.27$ (see Table \ref{tab:fuentes_ssfr}). These properties differ from those of  strongly lensed galaxies in the well-studied  PASSAGES (Planck All-Sky Survey to Analyze Gravitationally-lensed Extreme Starbursts) and South Pole Telescope (SPT) samples. The PASSAGES sample of 30 galaxies have median redshift of $\bar{z}\approx 2$ and $\rm \overline{SFR}\approx 1500\,M_\odot\,yr^{-1}$ \citep[e.g., ][]{kamieneski2024passages}; while the  81 SPT-selected galaxies have, typically higher redshifts and SFRs with median $\bar{z}\approx 3.9$ and $\rm \overline{SFR} \approx 2300\,M_\odot\,yr^{-1}$ \citep[e.g.,][]{reuter2020complete,liu2024detailed}. These systems, therefore, have been known as the most extreme starbursts at high redshifts. As illustrated in Figure~\ref{fig:sfr_mass_magnified}, our sample exhibits properties closely resembling the unlensed population of submillimeter galaxies (SMGs) from the AS2UDS survey \citep{dudzevivciute2020alma}, which is characterized by a median stellar mass of $\log(M_\star/\rm{M}_{\odot})\approx 11.1$.

To illustrate how our sample of 22 moderately/strongly lensed galaxies in the Frontier Fields compares with the SPT and PASSAGES samples, in Figure\,\ref{fig:sfr_mass_magnified} we plot their sSFR as a function of redshift. It is evident that the Frontier Fields sample exhibit lower sSFR  and redshifts than the SPT and PASSAGES samples. While the SPT and PASSAGES sample have opened a window into the star formation conditions of massive, starburst systems at $z\approx 2-7$ \citep[e.g.,][]{ma2015stellar,reuter2020complete}, the Frontier Field sample presented here can be used to zoom into galaxy evolution processes of more typical, main sequence galaxies at the peak epoch of star formation in the Universe ($z\approx 1-2$).

\section{Summary} \label{sec:summary}
Using the VLA C-band receivers centered at 6\,GHz, we generated the deepest (rms noise $\approx 1\,\mu$Jy\,beam$^{-1}$), high resolution ($0\farcs 82$) radio image to date of A\,2744 ---the third strongest lensing cluster from the six Hubble Frontier Fields. The  main data products and results derived from this work are the following.

\begin{itemize}

\item The radio source catalog contains 93 sources detected with a peak SNR $> 5$. Five of them are extended or multi-component sources, and 88 are cataloged as point-like sources. The total expected fraction of spurious sources in our radio catalog is 16\% (15 sources). However, the fraction of spurious sources in the subset of 48 VLA sources within the JWST coverage is $\approx 4\%$, suggesting that the reliability of our catalog is higher than expected from the negative source count analysis.\\

\item The 6\,GHz sample has a median effective radius of $0\farcs267$, corresponding to a $\sim 2.01$\,kpc at $z = 0.93$, and a median peak surface brightness of 15.6$\,\mu$Jy\,beam$^{-1}$. \\

\item We cross-matched our 6\,GHz radio source catalog with the UNCOVER \citep[][]{wang2023uncover,2024ApJS..270....7W} to get galaxy properties as: redshift, magnification, stellar mass, and NIR/optic SFRs. Within the region of overlapping footprints, 46 out of 48 radio sources have JWST and HST counterparts within a 0\farcs5 search radius. The remaining two sources lack optical/NIR counterparts and are likely spurious, although the possibility that they are dust-enshrouded galaxies cannot be ruled out.

\item The radio host galaxies are in the redshift range  $0.15 < z  \leq 3.55$ with a median of $z_{\rm{spec}} = 0.30^{+0.06}_{-0.02}$ and $z_{\rm{phot}} = 1.07^{+1.55}_{-0.68}$, with 11 of them being moderately magnified (i.e,. $1 \leq \mu \leq 9.27$). The  stellar masses span from $6.6 \times 10^{5}\,\rm{M}_{\odot}$ to $1.6 \times 10^{11}\,\rm{M}_{\odot}$. The median $\log(\mathrm{SFR})$ is $0.9^{+1.7}_{-0.2}$ from NIR/optical photometry, $2.2^{+0.4}_{-1.2}$ from 6\,GHz data, and $0.3^{+0.9}_{-0.1}\,\mathrm{M}_{\odot}\,\mathrm{yr}^{-1}$ from the $u$-band. \\

\item Within the subsample of 46 sources, we identified 7 AGN candidates based on X-ray luminosity ($L_{\rm{x}} > 10^{42}$ erg s$^{-1}$) and 2 sources exhibiting a radio excess relative to the IR-radio correlation, with one source overlapping both criteria. Additionally, a visual inspection of the 93 radio sources revealed one AGN with clear radio lobes. In total, we identified 9 AGN candidates, {representing a fraction of $\sim 20 \%$.} This result is broadly consistent with the predicted AGN fraction of $\sim 14 \%$ by \cite{mancuso2017galaxy} and \cite{bonaldi2019tiered}. \\

\item We computed dust un-obscured (rest-frame $u$-band) and obscured (radio) SFRs for 46 VLA sources with available redshift and  $u$-band flux densities. The radio-based SFRs are typically larger than those from $u$-band imaging. We also compare the 6\,GHz SFRs with those reported by \citet{2024ApJS..270....7W} using SED fitting of JWST/HST photometric data, revealing that the former are higher by a factor of 5 to 50. \\

\item None of the 23 LRDs at $z\approx6$ reported by \citet{gloudemans2025another} are detected in the 6\,GHz map. After stacking, we derive a 3$\sigma$ upper limit to the 6\,GHz radio luminosity of $4.1\times 10^{39}\,\rm erg\,s^{-1}$.\\

\item The 22 galaxies in the VLA Frontier Field survey that are moderately/strongly lensed ($\mu>2$) probe a sSFR regime that has been largely missed by existing samples of strongly lensed galaxies, like PASSAGES and SPT, facilitating spatially resolved  studies of star formation in more typical, main sequence galaxies at $z\approx1-2$.

\end{itemize}
\begin{acknowledgments}
We thank the anonymous referee for a careful and insightful review that significantly improved the quality and clarity of this manuscript.
E.A.O.  and E.F.-J.A. acknowledge
support from the  Program to Support Research and Technological Innovation Projects (PAPIIT; Projects IA102023 and IA104725) of the National Autonomous University of Mexico (UNAM), and from the Program ``Frontier Science”   (Project ID CF-2023-I-506)  of the National Council of Humanities, Sciences and Technologies (CONAHCyT). The National Radio Astronomy Observatory is a facility of the National Science Foundation operated under cooperative agreement by Associated Universities, Inc.
E.A.O thanks the Hubble Frontier Fields Project team (PI: J. Lotz) for their observations that contributed to this work.
Special thanks are extended to the ``Ultradeep NIRSpec and NIRCam ObserVations before the Epoch of Reionization'' (UNCOVER) project team (PIs: Ivo Labbe \& Rachel Bezanson) for providing the invaluable {\it HST}  and {\it JWST} data and observations that were incorporated into this study. 
IRS acknowledges the support of the Science and Technology Facilities Council (STFC) under grant ST/X001075/1.

This work is based on observations made with the NASA/ESA/CSA James Webb Space Telescope. The data were obtained from the Mikulski Archive for Space Telescopes at the Space Telescope Science Institute, which is operated by the Association of Universities for Research in Astronomy, Inc., under NASA contract NAS 5-03127 for JWST. These observations are associated with programs \href{https://www.stsci.edu/jwst/science-execution/program-information?id=1324}{ERS-1324}, \href{https://www.stsci.edu/jwst/science-execution/program-information?id=2561}{GO-2561}, and \href{https://www.stsci.edu/jwst/science-execution/program-information?id=2756}{DD-2756}.

The JWST data described here may be obtained from
\url{https://dx.doi.org/10.17909/kw3c-n857}.

\facilities{NSF’s Karl G. Jansky Very Large Array
(VLA), \textit{HST}, \textit{JWST}, and ALMA.}

\software{astropy \citep{robitaille2013astropy},  
          CASA \citep{bean2022casa}, 
          PyBDSF \citep{mohan2015pybdsf}}

\end{acknowledgments}

\section*{Data Availability}
The VLA Frontier Fields survey is a public legacy project,
and we make all our catalog and image products freely available at \url{https://science.nrao.edu/science/surveys/vla-ff}.

\appendix

\section{Tables and RGB images}

Here we present the tables with the positions, radio sizes, the flux densities of the 93 radio sources reported in this work. We report the redshifts, magnification factors, stellar masses, and star‐formation rates derived from UNCOVER \citep{wang2023uncover}, as well as the sSFR and other properties of the moderately/strongly lensed ($\mu > 2$) galaxies from the SPT and VLA surveys.  Additionally,  we show RGB images from the 46 sources with \textit{JWST + HST} counterparts, including the source N.54 and N.91 (see Figure \ref{fig:RGB_Images_2}). 

\startlongtable
} \\ \hline
20 & J001428.61-302270.9 & 0.61 & p & 2.46  & $10.36$ & $0.73\pm0.35$     \\
28 & J001426.07-302452.5 & 2.65 & p & 2.18  & $11.12$ & $2.52\pm1.99$   \\
29 & J001425.35-302550.3 & 2.90 & p & 2.26  & $10.13$ & $2.25\pm1.89$   \\
38 & J001421.68-302410.2 & 0.50 & s & 3.01  & $10.70$ & $1.05\pm0.02$    \\
46 & J001419.80-302370.6 & 2.93 & p & 4.60  & $10.00$ & $2.20\pm1.69$   \\
47 & J001419.51-302248.9 & 3.55 & p & 3.14  & $10.53$ & $2.48\pm2.02$   \\
51 & J001417.58-302300.6 & 1.22 & p & 2.48  & $09.81$ & $2.22\pm0.96$   \\
62 & J001413.92-302237.9 & 2.93 & p & 9.27  & $09.74$ & $2.28\pm1.68$   \\
85 & J001405.71-302217.3 & 2.12 & p & 2.06  & $10.38$ & $2.38\pm1.94$   \\
- & J041606.36-240451.2 & 0.74 & s & 3.03  & 10.36 & $1.11\pm 0.16$    \\
- & J041606.62-240527.8 & 1.90 & p & 2.26  & 10.03 & $1.75\pm0.18$     \\
- & J041611.67-240419.6 & 2.20 & p & 2.26  & 10.16 & $1.94\pm 0.15$    \\
- & J071725.85+374446.2 & 2.93 & p & 2.21  & 10.36  & $2.16\pm0.15$   \\
- & J071730.65+374443.1 & 1.01 & s & 2.84  & 10.43 & $0.88\pm0.16$    \\
- & J071733.14+374543.2 & 0.91 & s & 2.11  & 09.89 & $1.38\pm0.14$    \\
- & J071734.46+374432.2 & 1.14 & s & 5.84  & 09.42 & $1.56\pm0.12$    \\
- & J071735.22+374541.7 & 1.69 & s & 3.61  & 10.87 & $1.73\pm0.14$   \\
- & J071736.66+374506.4 & 1.13 & s & 6.45  & 09.48 & $1.01\pm0.15$     \\
- & J071740.55+374506.4 & 1.97 & p & 2.18  & 10.48 & $1.83\pm0.15$    \\
- & J114932.03+222439.3 & 1.28 & s & 2.11  & 10.19 & $2.12\pm0.19$    \\
- & J114934.46+222438.5 & 0.75 & s & 2.16  & 09.77 & $0.72\pm0.13$     \\ 
- & J114936.09+222424.4 & 1.64 & p & 3.13  & 10.71 & $1.67\pm0.15$    \\\hline
\end{tabular*}
\caption{The sample of 22 moderately/strongly lensed ($\mu \geq 2$) galaxies in the VLA Frontier Fields survey, including 12 galaxies in the MACSJ0416.1-2403, MACSJ0717.5+3745, and MACSJ1149.5+2223 \citep{heywood2021vla, jimenez2021vla} and 9 galaxies in A\,2744 reported in this work.$^*$p denotes the photometric redshift and s denotes spectroscopic redshift. }
\label{tab:fuentes_ssfr}
\end{table}
\FloatBarrier 

\begin{figure}[p]
    \centering
    \includegraphics[width=\textwidth,height=0.95\textheight,keepaspectratio]{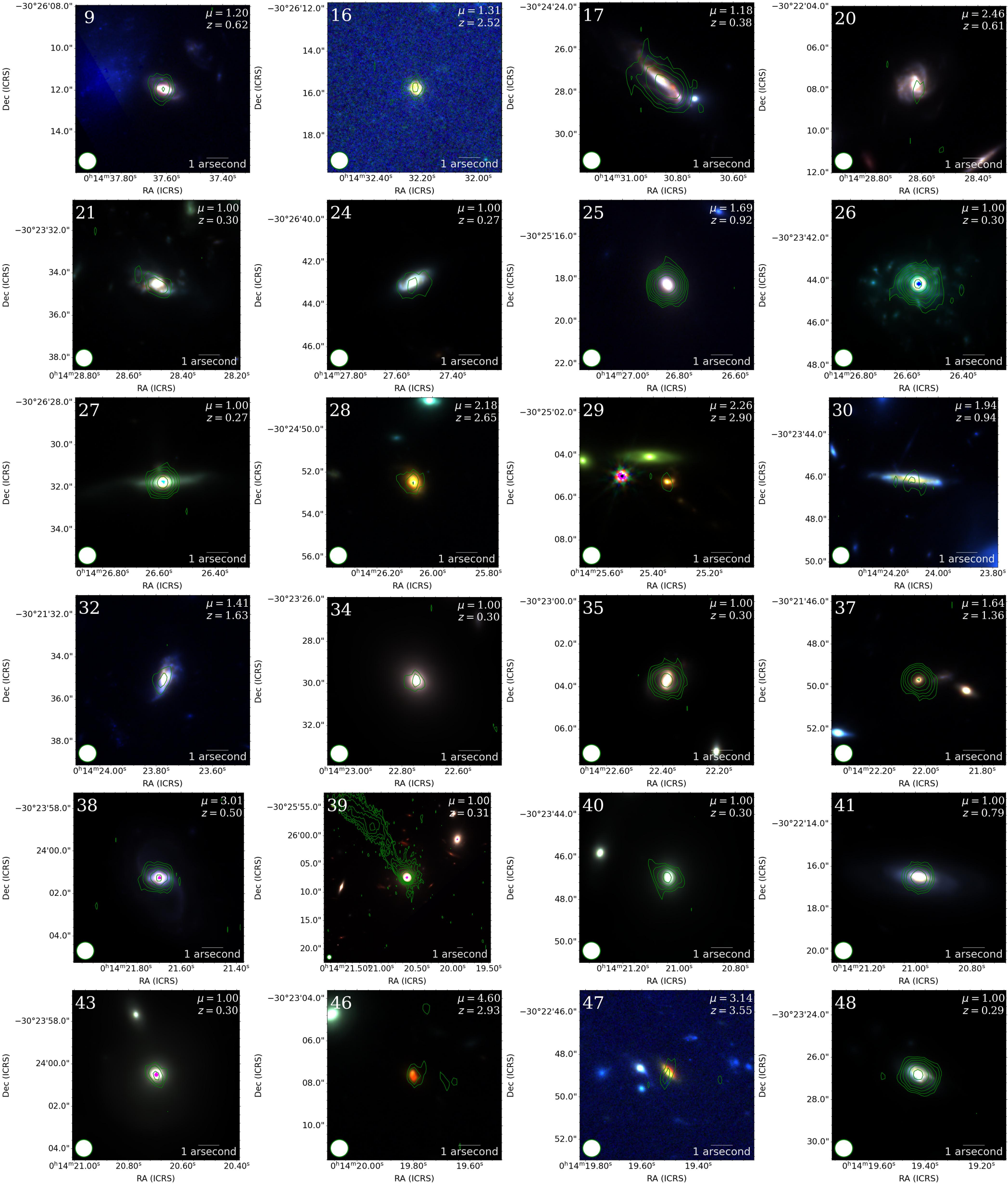}
\caption{}
\end{figure}

\clearpage 

\begin{figure}[p]
    \ContinuedFloat
    \centering
    \includegraphics[width=\textwidth,height=0.95\textheight,keepaspectratio]{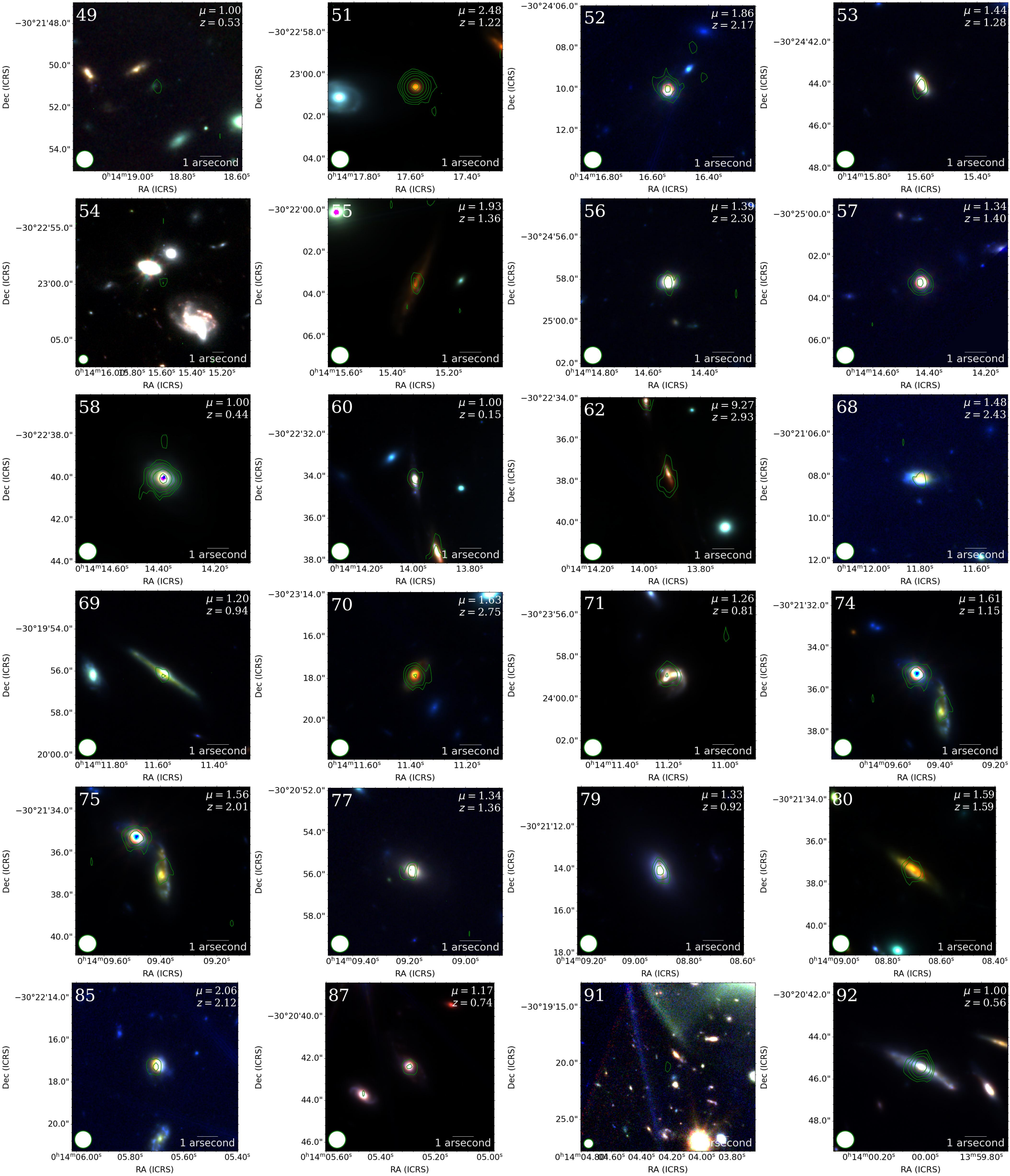}
    \caption{RGB NIRCam (R: F444W, G: F277W, B: F150W) images for the 46 radio sources detected at 6\,GHz with a \textit{HST}+\textit{JWST} counterpart, including source N.54 and N.91. The green contours represent the 3$\sigma$, 5$\sigma$, 8$\sigma$, 13$\sigma$, 21$\sigma$, 34$\sigma$ and 55$\sigma$ significance levels of the VLA images. The green circle in the bottom left represents the VLA synthesized beam ($0\farcs$82).}
    \label{fig:RGB_Images_2}
\end{figure}

\FloatBarrier 

\bibliographystyle{aasjournal}
\bibliography{sample7}

\begin{thebibliography}{}
\expandafter\ifx\csname natexlab\endcsname\relax\def\natexlab#1{#1}\fi
\providecommand{\url}[1]{\href{#1}{#1}}
\providecommand{\dodoi}[1]{doi:~\href{http://doi.org/#1}{\nolinkurl{#1}}}
\providecommand{\doeprint}[1]{\href{http://ascl.net/#1}{\nolinkurl{http://ascl.net/#1}}}
\providecommand{\doarXiv}[1]{\href{https://arxiv.org/abs/#1}{\nolinkurl{https://arxiv.org/abs/#1}}}

\bibitem[{Algera {et~al.}(2020)Algera, Smail, Dudzevi{\v{c}}i{\=u}t{\.e}, Swinbank, Stach, Hodge, Thomson, Almaini, Arumugam, Blain, {et~al.}}]{algera2020alma}
Algera, H., Smail, I., Dudzevi{\v{c}}i{\=u}t{\.e}, U., {et~al.} 2020, \href{https://iopscience.iop.org/article/10.3847/1538-4357/abb77b/meta}{ApJ}, 903, 138

\bibitem[{Ananna {et~al.}(2024)Ananna, Bogd{\'a}n, Kov{\'a}cs, Natarajan, \& Hickox}]{ananna2024x}
Ananna, T.~T., Bogd{\'a}n, {\'A}., Kov{\'a}cs, O.~E., Natarajan, P., \& Hickox, R.~C. 2024, \href{https://iopscience.iop.org/article/10.3847/2041-8213/ad5669/meta}{ApJL}, 969, L18

\bibitem[{{Arango-Toro} {et~al.}(2023){Arango-Toro}, {Ciesla}, {Ilbert}, {Magnelli}, {Jim{\'e}nez-Andrade}, \& {Buat}}]{2023A&A...675A.126A}
{Arango-Toro}, R.~C., {Ciesla}, L., {Ilbert}, O., {et~al.} 2023, \aap, 675, A126, \dodoi{10.1051/0004-6361/202345848}

\bibitem[{Barro {et~al.}(2024)Barro, P{\'e}rez-Gonz{\'a}lez, Kocevski, McGrath, Trump, Simons, Somerville, Yung, Haro, Akins, {et~al.}}]{barro2024extremely}
Barro, G., P{\'e}rez-Gonz{\'a}lez, P.~G., Kocevski, D.~D., {et~al.} 2024, \href{https://iopscience.iop.org/article/10.3847/1538-4357/ad167e/meta}{ApJ}, 963, 128

\bibitem[{Bean {et~al.}(2022)Bean, Bhatnagar, Castro, Meyer, Emonts, Garcia, Garwood, Golap, Villalba, Harris, {et~al.}}]{bean2022casa}
Bean, B., Bhatnagar, S., Castro, S., {et~al.} 2022, \href{https://iopscience.iop.org/article/10.1088/1538-3873/ac9642/meta}{PASP}, 134, 114501

\bibitem[{Bezanson {et~al.}(2024)Bezanson, Labbe, Whitaker, Leja, Price, Franx, Brammer, Marchesini, Zitrin, Wang, {et~al.}}]{bezanson2024jwst}
Bezanson, R., Labbe, I., Whitaker, K.~E., {et~al.} 2024, \href{https://iopscience.iop.org/article/10.3847/1538-4357/ad66cf/meta}{ApJ}, 974, 92

\bibitem[{Blanton {et~al.}(2003)Blanton, Brinkmann, Csabai, Doi, Eisenstein, Fukugita, Gunn, Hogg, \& Schlegel}]{blanton2003estimating}
Blanton, M.~R., Brinkmann, J., Csabai, I., {et~al.} 2003, \href{https://iopscience.iop.org/article/10.1086/342935/meta}{AJ}, 125, 2348

\bibitem[{Bonaldi {et~al.}(2019)Bonaldi, Bonato, Galluzzi, Harrison, Massardi, Kay, De~Zotti, \& Brown}]{bonaldi2019tiered}
Bonaldi, A., Bonato, M., Galluzzi, V., {et~al.} 2019, \href{https://academic.oup.com/mnras/article/482/1/2/5108200}{MNRAS}, 482, 2

\bibitem[{Bouwens {et~al.}(2020)Bouwens, Gonz{\'a}lez-L{\'o}pez, Aravena, Decarli, Novak, Stefanon, Walter, Boogaard, Carilli, Dudzevi{\v{c}}i{\=u}t{\.e}, {et~al.}}]{bouwens2020alma}
Bouwens, R., Gonz{\'a}lez-L{\'o}pez, J., Aravena, M., {et~al.} 2020, \href{https://iopscience.iop.org/article/10.3847/1538-4357/abb830/meta}{ApJ}, 902, 112

\bibitem[{Bruzual \& Charlot(2003)}]{bruzual2003stellar}
Bruzual, G., \& Charlot, S. 2003, \href{https://doi.org/10.1046/j.1365-8711.2003.06897.x}{MNRAS}, 344, 1000

\bibitem[{Calzetti(2001)}]{calzetti2001dust}
Calzetti, D. 2001, \href{https://iopscience.iop.org/article/10.1086/324269/meta}{PASP}, 113, 1449

\bibitem[{Chapman {et~al.}(2004)Chapman, Smail, Blain, \& Ivison}]{chapman2004population}
Chapman, S.~C., Smail, I., Blain, A., \& Ivison, R. 2004, \href{https://iopscience.iop.org/article/10.1086/423833/meta}{ApJ}, 614, 671

\bibitem[{Chen {et~al.}(2022)Chen, Kelly, Treu, Williams, Broadhurst, Castellano, Diego, Filippenko, Glazebrook, Koekemoer, {et~al.}}]{chen2022imaging}
Chen, W., Kelly, P., Treu, T.~L., {et~al.} 2022, JWST Proposal. Cycle 1, 2756

\bibitem[{Choi {et~al.}(2016)Choi, Dotter, Conroy, Cantiello, Paxton, \& Johnson}]{choi2016mesa}
Choi, J., Dotter, A., Conroy, C., {et~al.} 2016, \href{https://iopscience.iop.org/article/10.3847/0004-637X/823/2/102/meta}{ApJ}, 823, 102

\bibitem[{Condon(1992)}]{condon1992radio}
Condon. 1992, \href{https://adsabs.harvard.edu/full/1992ARA%26A..30..575C�Ü�Ü}{ARA\&A}, 30, 575

\bibitem[{Conroy \& Gunn(2010)}]{conroy2010propagation}
Conroy, C., \& Gunn, J.~E. 2010, \href{https://iopscience.iop.org/article/10.1088/0004-637X/712/2/833/meta}{ApJ}, 712, 833

\bibitem[{Davidson-Pilon(2019)}]{Davidson-Pilon2019}
Davidson-Pilon, C. 2019, \href{https://joss.theoj.org/papers/10.21105/joss.01317}{JOSS}, 4, 1317

\bibitem[{Delhaize {et~al.}(2017)Delhaize, Smol{\v{c}}i{\'c}, Delvecchio, Novak, Sargent, Baran, Magnelli, Zamorani, Schinnerer, Murphy, {et~al.}}]{delhaize2017vla}
Delhaize, J., Smol{\v{c}}i{\'c}, V., Delvecchio, I., {et~al.} 2017, \href{https://www.aanda.org/component/article?access=doi&doi=10.1051/0004-6361/201629430}{A\&A}, 602, A4

\bibitem[{Dotter(2016)}]{dotter2016mesa}
Dotter, A. 2016, \href{https://iopscience.iop.org/article/10.3847/0067-0049/222/1/8/meta}{ApJ}, 222, 8

\bibitem[{Draine \& Li(2007)}]{draine2007infrared}
Draine, B.~T., \& Li, A. 2007, \href{https://iopscience.iop.org/article/10.1086/511055/meta}{ApJ}, 657, 810

\bibitem[{Dudzevi{\v{c}}i{\=u}t{\.e} {et~al.}(2020)Dudzevi{\v{c}}i{\=u}t{\.e}, Smail, Swinbank, Stach, Almaini, da~Cunha, An, Arumugam, Birkin, Blain, {et~al.}}]{dudzevivciute2020alma}
Dudzevi{\v{c}}i{\=u}t{\.e}, U., Smail, I., Swinbank, A., {et~al.} 2020, \href{https://academic.oup.com/mnras/article/494/3/3828/5815090}{MNRAS}, 494, 3828

\bibitem[{Eales {et~al.}(2010)Eales, Dunne, Clements, Cooray, De~Zotti, Dye, Ivison, Jarvis, Lagache, Maddox, {et~al.}}]{eales2010herschel}
Eales, S., Dunne, L., Clements, D., {et~al.} 2010, \href{https://iopscience.iop.org/article/10.1086/653086/meta}{PASP}, 122, 499

\bibitem[{Evans \& Civano(2018)}]{evans2018chandra}
Evans, I.~N., \& Civano, F. 2018, \href{https://academic.oup.com/astrogeo/article-abstract/59/2/ASTROG/4935787?redirectedFrom=PDF#page=16}{A\&G}, 59, 2

\bibitem[{Feigelson \& Nelson(1985)}]{feigelson1985statistical}
Feigelson, E.~D., \& Nelson, P.~I. 1985, \href{https://adsabs.harvard.edu/full/1985ApJ...293..192F}{ApJ}, 293, 192

\bibitem[{Fujimoto {et~al.}(2023)Fujimoto, Bezanson, Labbe, Brammer, Price, Wang, Weaver, Fudamoto, Oesch, Williams, {et~al.}}]{fujimoto2023dualz}
Fujimoto, S., Bezanson, R., Labbe, I., {et~al.} 2023, \href{https://arxiv.org/abs/2309.07834}{arXiv preprint arXiv:2309.07834}

\bibitem[{Gillman {et~al.}(2024)Gillman, Smail, Gullberg, Swinbank, Vijayan, Lee, Brammer, Dudzevi{\v{c}}i{\=u}t{\.e}, Greve, Almaini, {et~al.}}]{gillman2024structure}
Gillman, S., Smail, I., Gullberg, B., {et~al.} 2024, Astronomy \& Astrophysics, 691, A299

\bibitem[{Gloudemans {et~al.}(2025)Gloudemans, Duncan, Eilers, Farina, Harikane, Inayoshi, Lambrides, \& Vardoulaki}]{gloudemans2025another}
Gloudemans, A.~J., Duncan, K.~J., Eilers, A.-C., {et~al.} 2025, \href{10.3847/1538-4357/adddb9}{AAS}, 986, 130

\bibitem[{G{\'o}mez-Guijarro {et~al.}(2022)G{\'o}mez-Guijarro, Elbaz, Xiao, B{\'e}thermin, Franco, Magnelli, Daddi, Dickinson, Demarco, Inami, {et~al.}}]{gomez2022goods}
G{\'o}mez-Guijarro, C., Elbaz, D., Xiao, M., {et~al.} 2022, \href{https://www.aanda.org/articles/aa/abs/2022/02/aa41615-21/aa41615-21.html}{A\&A}, 658, A43

\bibitem[{Gonz{\'a}lez-L{\'o}pez {et~al.}(2020)Gonz{\'a}lez-L{\'o}pez, Novak, Decarli, Walter, Aravena, Carilli, Boogaard, Popping, Weiss, Assef, {et~al.}}]{gonzalez2020alma}
Gonz{\'a}lez-L{\'o}pez, J., Novak, M., Decarli, R., {et~al.} 2020, \href{https://iopscience.iop.org/article/10.3847/1538-4357/ab765b/meta}{ApJ}, 897, 91

\bibitem[{{González-López} {et~al.}(2017){González-López}, {Bauer, F. E.}, {Romero-Cañizales, C.}, {Kneissl, R.}, {Villard, E.}, {Carvajal, R.}, {Kim, S.}, {Laporte, N.}, {Anguita, T.}, {Aravena, M.}, {Bouwens, R. J.}, {Bradley, L.}, {Carrasco, M.}, {Demarco, R.}, {Ford, H.}, {Ibar, E.}, {Infante, L.}, {Messias, H.}, {Muñoz Arancibia, A. M.}, {Nagar, N.}, {Padilla, N.}, {Treister, E.}, {Troncoso, P.}, \& {Zitrin, A.}}]{gonzales2017}
{González-López}, {Bauer, F. E.}, {Romero-Cañizales, C.}, {et~al.} 2017, \href{https://doi.org/10.1051/0004-6361/201628806}{A\&A}, 597, A41

\bibitem[{Greene {et~al.}(2024)Greene, Labbe, Goulding, Furtak, Chemerynska, Kokorev, Dayal, Volonteri, Williams, Wang, {et~al.}}]{greene2024uncover}
Greene, J.~E., Labbe, I., Goulding, A.~D., {et~al.} 2024, \href{https://iopscience.iop.org/article/10.3847/1538-4357/ad1e5f/meta}{ApJ}, 964, 39

\bibitem[{Harikane {et~al.}(2023)Harikane, Zhang, Nakajima, Ouchi, Isobe, Ono, Hatano, Xu, \& Umeda}]{harikane2023jwst}
Harikane, Y., Zhang, Y., Nakajima, K., {et~al.} 2023, \href{https://iopscience.iop.org/article/10.3847/1538-4357/ad029e/meta}{ApJ}, 959, 39

\bibitem[{Helou {et~al.}(1985)Helou, Soifer, \& Rowan-Robinson}]{helou1985thermal}
Helou, G., Soifer, B., \& Rowan-Robinson, M. 1985, \href{https://adsabs.harvard.edu/full/record/seri/ApJ../0298/1985ApJ...298L...7H.html}{ApJ}, 298, L7

\bibitem[{Heywood {et~al.}(2021)Heywood, Murphy, Jim{\'e}nez-Andrade, Armus, Cotton, DeCoursey, Dickinson, Lazio, Momjian, Penner, {et~al.}}]{heywood2021vla}
Heywood, I., Murphy, E., Jim{\'e}nez-Andrade, E., {et~al.} 2021, \href{https://iopscience.iop.org/article/10.3847/1538-4357/abdf61/meta}{ApJ}, 910, 105

\bibitem[{Hopkins {et~al.}(2003)Hopkins, Miller, Nichol, Connolly, Bernardi, G{\'o}mez, Goto, Tremonti, Brinkmann, Ivezi{\'c}, {et~al.}}]{hopkins2003star}
Hopkins, A.~M., Miller, C., Nichol, R., {et~al.} 2003, \href{https://iopscience.iop.org/article/10.1086/379608/meta}{ApJ}, 599, 971

\bibitem[{Isobe {et~al.}(1990)Isobe, Feigelson, Akritas, \& Babu}]{isobe1990linear}
Isobe, T., Feigelson, E.~D., Akritas, M.~G., \& Babu, G.~J. 1990, \href{https://adsabs.harvard.edu/full/1990ApJ...364..104I}{ApJ}, 364, 104

\bibitem[{Jim{\'e}nez-Andrade {et~al.}(2021)Jim{\'e}nez-Andrade, Murphy, Heywood, Smail, Penner, Momjian, Dickinson, Armus, \& Lazio}]{jimenez2021vla}
Jim{\'e}nez-Andrade, E., Murphy, E., Heywood, I., {et~al.} 2021, \href{https://iopscience.iop.org/article/10.3847/1538-4357/abe876/meta}{ApJ}, 910, 106

\bibitem[{Johnson {et~al.}(2021)Johnson, Leja, Conroy, \& Speagle}]{johnson2021stellar}
Johnson, B.~D., Leja, J., Conroy, C., \& Speagle, J.~S. 2021, \href{https://iopscience.iop.org/article/10.3847/1538-4365/abef67/meta}{ApJ}, 254, 22

\bibitem[{Kamieneski {et~al.}(2024)Kamieneski, Yun, Harrington, Lowenthal, Wang, Frye, Jim{\'e}nez-Andrade, Vishwas, Cooper, Pascale, {et~al.}}]{kamieneski2024passages}
Kamieneski, P.~S., Yun, M.~S., Harrington, K.~C., {et~al.} 2024, \href{https://iopscience.iop.org/article/10.3847/1538-4357/acf930/meta}{ApJ}, 961, 2

\bibitem[{Kaplan \& Meier(1958)}]{kaplan1958nonparametric}
Kaplan, E.~L., \& Meier, P. 1958, \href{https://www.tandfonline.com/doi/abs/10.1080/01621459.1958.10501452}{JASA}, 53, 457

\bibitem[{Klein {et~al.}(2018)Klein, Lisenfeld, \& Verley}]{klein2018radio}
Klein, U., Lisenfeld, U., \& Verley, S. 2018, \href{https://www.aanda.org/articles/aa/abs/2018/03/aa31673-17/aa31673-17.html}{A\&A}, 611, A55

\bibitem[{Kocevski {et~al.}(2023)Kocevski, Onoue, Inayoshi, Trump, Haro, Grazian, Dickinson, Finkelstein, Kartaltepe, Hirschmann, {et~al.}}]{kocevski2023hidden}
Kocevski, D.~D., Onoue, M., Inayoshi, K., {et~al.} 2023, \href{https://iopscience.iop.org/journal/2041-8205}{ApJL}, 954, L4

\bibitem[{Kocevski {et~al.}(2024)Kocevski, Finkelstein, Barro, Taylor, Calabr{\`o}, Laloux, Buchner, Trump, Leung, Yang, {et~al.}}]{kocevski2024rise}
Kocevski, D.~D., Finkelstein, S.~L., Barro, G., {et~al.} 2024, \href{https://arxiv.org/abs/2404.03576}{arXiv preprint arXiv:2404.03576}

\bibitem[{Labb{\'e} {et~al.}(2023)Labb{\'e}, van Dokkum, Nelson, Bezanson, Suess, Leja, Brammer, Whitaker, Mathews, Stefanon, {et~al.}}]{labbe2023population}
Labb{\'e}, I., van Dokkum, P., Nelson, E., {et~al.} 2023, \href{https://www.nature.com/articles/s41586-023-05786-2}{Nature}, 616, 266

\bibitem[{Labb{\'e} {et~al.}(2024)Labb{\'e}, Greene, Bezanson, Fujimoto, Furtak, Goulding, Matthee, Naidu, Oesch, Atek, {et~al.}}]{labbe2024uncover}
Labb{\'e}, I., Greene, J.~E., Bezanson, R., {et~al.} 2024, \href{https://iopscience.iop.org/article/10.3847/1538-4357/ad3551/meta}{ApJ}, 978, 92

\bibitem[{{Lacki} {et~al.}(2010){Lacki}, {Thompson}, \& {Quataert}}]{2010ApJ...717....1L}
{Lacki}, B.~C., {Thompson}, T.~A., \& {Quataert}, E. 2010, \apj, 717, 1, \dodoi{10.1088/0004-637X/717/1/1}

\bibitem[{Laigle {et~al.}(2016)Laigle, McCracken, Ilbert, Hsieh, Davidzon, Capak, Hasinger, Silverman, Pichon, Coupon, {et~al.}}]{laigle2016cosmos2015}
Laigle, C., McCracken, H.~J., Ilbert, O., {et~al.} 2016, \href{https://iopscience.iop.org/article/10.3847/0067-0049/224/2/24/meta}{ApJ}, 224, 24

\bibitem[{{Laporte} {et~al.}(2017){Laporte}, {Bauer, F. E.}, {Troncoso-Iribarren, P.}, {Huang, X.}, {González-López, J.}, {Kim, S.}, {Anguita, T.}, {Aravena, M.}, {Barrientos, L. F.}, {Bouwens, R.}, {Bradley, L.}, {Brammer, G.}, {Carrasco, M.}, {Carvajal, R.}, {Coe, D.}, {Demarco, R.}, {Ellis, R. S.}, {Ford, H.}, {Francke, H.}, {Ibar, E.}, {Infante, L.}, {Kneissl, R.}, {Koekemoer, A. M.}, {Messias, H.}, {Muñoz Arancibia, A.}, {Nagar, N.}, {Padilla, N.}, {Pelló, R.}, {Postman, M.}, {Quénard, D.}, {Romero-Cañizales, C.}, {Treister, E.}, {Villard, E.}, {Zheng, W.}, \& {Zitrin, A.}}]{Laporte2017}
{Laporte}, {Bauer, F. E.}, {Troncoso-Iribarren, P.}, {et~al.} 2017, \href{https://doi.org/10.1051/0004-6361/201730628}{A\&A}, 604, A132

\bibitem[{Leslie {et~al.}(2020)Leslie, Schinnerer, Liu, Magnelli, Algera, Karim, Davidzon, Gozaliasl, Jim{\'e}nez-Andrade, Lang, {et~al.}}]{leslie2020vla}
Leslie, S.~K., Schinnerer, E., Liu, D., {et~al.} 2020, \href{https://iopscience.iop.org/article/10.3847/1538-4357/aba044/meta}{ApJ}, 899, 58

\bibitem[{Liu {et~al.}(2024)Liu, F{\"o}rster~Schreiber, Harrington, Lee, Kamieneski, Davies, Lutz, Renzini, Wuyts, Tacconi, {et~al.}}]{liu2024detailed}
Liu, D., F{\"o}rster~Schreiber, N.~M., Harrington, K.~C., {et~al.} 2024, \href{https://www.nature.com/articles/s41550-024-02296-7}{Nat. Astron.}, 8, 1181

\bibitem[{{Lo Faro} {et~al.}(2015){Lo Faro}, {Silva}, {Franceschini}, {Miller}, \& {Efstathiou}}]{2015MNRAS.447.3442L}
{Lo Faro}, B., {Silva}, L., {Franceschini}, A., {Miller}, N., \& {Efstathiou}, A. 2015, \mnras, 447, 3442, \dodoi{10.1093/mnras/stu2593}

\bibitem[{Lotz {et~al.}(2017)Lotz, Koekemoer, Coe, Grogin, Capak, Mack, Anderson, Avila, Barker, Borncamp, {et~al.}}]{lotz2017frontier}
Lotz, J.~e., Koekemoer, A., Coe, D., {et~al.} 2017, \href{https://iopscience.iop.org/article/10.3847/1538-4357/837/1/97/meta}{ApJ}, 837, 97

\bibitem[{Lutz {et~al.}(2011)Lutz, Poglitsch, Altieri, Andreani, Aussel, Berta, Bongiovanni, Brisbin, Cava, Cepa, {et~al.}}]{lutz2011pacs}
Lutz, D., Poglitsch, A., Altieri, B., {et~al.} 2011, \href{https://doi.org/10.1051/0004-6361/201117107}{A\& A}, 532, A90

\bibitem[{Ma {et~al.}(2015)Ma, Gonzalez, Spilker, Strandet, Ashby, Aravena, B{\'e}thermin, Bothwell, De~Breuck, Brodwin, {et~al.}}]{ma2015stellar}
Ma, J., Gonzalez, A.~H., Spilker, J., {et~al.} 2015, \href{https://iopscience.iop.org/article/10.1088/0004-637X/812/1/88/meta}{ApJ}, 812, 88

\bibitem[{Madau(1995)}]{madau1995radiative}
Madau, P. 1995, \href{https://adsabs.harvard.edu/full/record/seri/ApJ../0441/1995ApJ...441...18M.html}{ApJ}, 441, 18

\bibitem[{Magnelli {et~al.}(2015)Magnelli, Ivison, Lutz, Valtchanov, Farrah, Berta, Bertoldi, Bock, Cooray, Ibar, {et~al.}}]{magnelli2015far}
Magnelli, B., Ivison, R., Lutz, D., {et~al.} 2015, \href{https://www.aanda.org/articles/aa/abs/2015/01/aa24937-14/aa24937-14.html}{A\&A}, 573, A45

\bibitem[{Maiolino {et~al.}(2025)Maiolino, Risaliti, Signorini, Trefoloni, Juod{\v{z}}balis, Scholtz, {\"U}bler, D’Eugenio, Carniani, Fabian, {et~al.}}]{maiolino2025jwst}
Maiolino, R., Risaliti, G., Signorini, M., {et~al.} 2025, \href{https://academic.oup.com/mnras/article/538/3/1921/8045610}{MNRAS}, 538, 1921

\bibitem[{Mancuso {et~al.}(2017)Mancuso, Lapi, Prandoni, Obi, Gonzalez-Nuevo, Perrotta, Bressan, Celotti, \& Danese}]{mancuso2017galaxy}
Mancuso, C., Lapi, A., Prandoni, I., {et~al.} 2017, \href{https://iopscience.iop.org/article/10.3847/1538-4357/aa745d/meta}{ApJ}, 842, 95

\bibitem[{Matthee {et~al.}(2024)Matthee, Naidu, Brammer, Chisholm, Eilers, Goulding, Greene, Kashino, Labbe, Lilly, {et~al.}}]{matthee2024little}
Matthee, J., Naidu, R.~P., Brammer, G., {et~al.} 2024, \href{https://iopscience.iop.org/article/10.3847/1538-4357/ad2345/meta}{ApJ}, 963, 129

\bibitem[{Mazzolari {et~al.}(2024)Mazzolari, Gilli, Maiolino, Prandoni, Delvecchio, Norman, Jimenez-Andrade, Belladitta, Vito, Momjian, {et~al.}}]{mazzolari2024radio}
Mazzolari, G., Gilli, R., Maiolino, R., {et~al.} 2024, \href{https://arxiv.org/abs/2412.04224}{arXiv preprint arXiv:2412.04224}

\bibitem[{Mohan \& Rafferty(2015)}]{mohan2015pybdsf}
Mohan, N., \& Rafferty, D. 2015, \href{https://ui.adsabs.harvard.edu/abs/2015ascl.soft02007M%2F/abstract}{ASCL}, 1502

\bibitem[{Moretti {et~al.}(2022)Moretti, Radovich, Poggianti, Vulcani, Gullieuszik, Werle, Bellhouse, Bacchini, Fritz, Soucail, {et~al.}}]{moretti2022observing}
Moretti, A., Radovich, M., Poggianti, B.~M., {et~al.} 2022, \href{https://iopscience.iop.org/article/10.3847/1538-4357/ac36c7/meta}{ApJ}, 925, 4

\bibitem[{Murphy {et~al.}(2008)Murphy, Helou, Kenney, Armus, \& Braun}]{murphy2008connecting}
Murphy, E., Helou, G., Kenney, J., Armus, L., \& Braun, R. 2008, \href{https://iopscience.iop.org/article/10.1086/587123/meta}{ApJ}, 678, 828

\bibitem[{Murphy(2009)}]{murphy2009far}
Murphy, E.~J. 2009, \href{https://iopscience.iop.org/article/10.1088/0004-637X/706/1/482/meta}{ApJ}, 706, 482

\bibitem[{Murphy {et~al.}(2017)Murphy, Momjian, Condon, Chary, Dickinson, Inami, Taylor, \& Weiner}]{murphy2017goods}
Murphy, E.~J., Momjian, E., Condon, J.~J., {et~al.} 2017, \href{https://iopscience.iop.org/article/10.3847/1538-4357/aa62fd/meta}{ApJ}, 839, 35

\bibitem[{Murphy {et~al.}(2006)Murphy, Helou, Braun, Kenney, Armus, Calzetti, Draine, Kennicutt~Jr, Roussel, Walter, {et~al.}}]{murphy2006effect}
Murphy, E.~J., Helou, G., Braun, R., {et~al.} 2006, \href{https://iopscience.iop.org/article/10.1086/509722/meta}{ApJ}, 651, L111

\bibitem[{Nguyen {et~al.}(2010)Nguyen, Schulz, Levenson, Amblard, Arumugam, Aussel, Babbedge, Blain, Bock, Boselli, {et~al.}}]{nguyen2010hermes}
Nguyen, H., Schulz, B., Levenson, L., {et~al.} 2010, \href{https://doi.org/10.1051/0004-6361/201014680}{A\&A}, 518, L5

\bibitem[{Olowin(1988)}]{olowin1988x}
Olowin, R. 1988, \href{https://iopscience.iop.org/article/10.1086/132333/meta}{PASP}, 100, 1354

\bibitem[{Paul {et~al.}(2019)Paul, Salunkhe, Datta, \& Intema}]{paul2019low}
Paul, S., Salunkhe, S., Datta, A., \& Intema, H.~T. 2019, \href{https://academic.oup.com/mnras/article/489/1/446/5553380}{MNRAS}, 489, 446

\bibitem[{Pearce {et~al.}(2017)Pearce, Van~Weeren, Andrade-Santos, Jones, Forman, Br{\"u}ggen, Bulbul, Clarke, Kraft, Medezinski, {et~al.}}]{pearce2017vla}
Pearce, C., Van~Weeren, R., Andrade-Santos, F., {et~al.} 2017, \href{https://iopscience.iop.org/article/10.3847/1538-4357/aa7e2f/meta}{ApJ}, 845, 81

\bibitem[{Perger {et~al.}(2025)Perger, Fogasy, Frey, \& Gab{\'a}nyi}]{perger2025deep}
Perger, K., Fogasy, J., Frey, S., \& Gab{\'a}nyi, K. 2025, \href{https://www.aanda.org/articles/aa/abs/2025/01/aa52422-24/aa52422-24.html}{A\&A}, 693, L2

\bibitem[{Radcliffe {et~al.}(2021)Radcliffe, Barthel, Thomson, Garrett, Beswick, \& Muxlow}]{radcliffe2021nowhere}
Radcliffe, J.~F., Barthel, P., Thomson, A., {et~al.} 2021, \href{https://www.aanda.org/articles/aa/abs/2021/05/aa38591-20/aa38591-20.html}{A\&A}, 649, A27

\bibitem[{Rahaman {et~al.}(2021)Rahaman, Raja, Datta, Burns, Alden, \& Rapetti}]{rahaman2021x}
Rahaman, M., Raja, R., Datta, A., {et~al.} 2021, \href{https://academic.oup.com/mnras/article/505/1/480/6258440}{MNRAS}, 505, 480

\bibitem[{Rawle {et~al.}(2016)Rawle, Altieri, Egami, P{\'e}rez-Gonz{\'a}lez, Boone, Clement, Ivison, Richard, Rujopakarn, Valtchanov, {et~al.}}]{rawle2016complete}
Rawle, T., Altieri, B., Egami, E., {et~al.} 2016, \href{https://academic.oup.com/mnras/article/459/2/1626/2595091}{MNRAS}, 459, 1626

\bibitem[{Reuter {et~al.}(2020)Reuter, Vieira, Spilker, Weiss, Aravena, Archipley, B{\'e}thermin, Chapman, De~Breuck, Dong, {et~al.}}]{reuter2020complete}
Reuter, C., Vieira, J., Spilker, J., {et~al.} 2020, \href{https://iopscience.iop.org/article/10.3847/1538-4357/abb599/meta}{ApJ}, 902, 78

\bibitem[{Robitaille {et~al.}(2013)Robitaille, Tollerud, Greenfield, Droettboom, Bray, Aldcroft, Davis, Ginsburg, Price-Whelan, Kerzendorf, {et~al.}}]{robitaille2013astropy}
Robitaille, T.~P., Tollerud, E.~J., Greenfield, P., {et~al.} 2013, \href{https://www.aanda.org/articles/aa/abs/2013/10/aa22068-13/aa22068-13.html}{A\&A}, 558, A33

\bibitem[{Smail {et~al.}(1991)Smail, Ellis, Fitchett, N{\o}rgaard-Nielsen, Hansen, \& J{\o}rgensen}]{smail1991statistically}
Smail, I., Ellis, R., Fitchett, M., {et~al.} 1991, \href{https://academic.oup.com/mnras/article/252/1/19/1284329}{MNRAS}, 252, 19

\bibitem[{Smol{\v{c}}i{\'c} {et~al.}(2017)Smol{\v{c}}i{\'c}, Novak, Bondi, Ciliegi, Mooley, Schinnerer, Zamorani, Navarrete, Bourke, Karim, {et~al.}}]{smolvcic2017vla}
Smol{\v{c}}i{\'c}, V., Novak, M., Bondi, M., {et~al.} 2017, \href{https://www.aanda.org/articles/aa/abs/2017/06/aa28704-16/aa28704-16.html}{A\&A}, 602, A1

\bibitem[{Steinhardt {et~al.}(2020)Steinhardt, Jauzac, Acebron, Atek, Capak, Davidzon, Eckert, Harvey, Koekemoer, Lagos, {et~al.}}]{steinhardt2020buffalo}
Steinhardt, C.~L., Jauzac, M., Acebron, A., {et~al.} 2020, \href{https://iopscience.iop.org/article/10.3847/1538-4365/ab75ed/meta}{ApJ}, 247, 64

\bibitem[{Treu {et~al.}(2022)Treu, Roberts-Borsani, Bradac, Brammer, Fontana, Henry, Mason, Morishita, Pentericci, Wang, {et~al.}}]{treu2022glass}
Treu, T., Roberts-Borsani, G., Bradac, M., {et~al.} 2022, \href{https://iopscience.iop.org/article/10.3847/1538-4357/ac8158/meta}{ApJ}, 935, 110

\bibitem[{Van~Weeren {et~al.}(2016)Van~Weeren, Ogrean, Jones, Forman, Andrade-Santos, Bonafede, Br{\"u}ggen, Bulbul, Clarke, Churazov, {et~al.}}]{van2016discovery}
Van~Weeren, R., Ogrean, G., Jones, C., {et~al.} 2016, \href{https://iopscience.iop.org/article/10.3847/0004-637X/817/2/98/meta}{ApJ}, 817, 98

\bibitem[{Wang {et~al.}(2023)Wang, Leja, Labb{\'e}, Bezanson, Whitaker, Brammer, Furtak, Weaver, Price, Zitrin, {et~al.}}]{wang2023uncover}
Wang, B., Leja, J., Labb{\'e}, I., {et~al.} 2023, \href{https://iopscience.iop.org/article/10.3847/1538-4365/ad0846/meta}{ApJ}, 270, 12

\bibitem[{Wang {et~al.}(2025)Wang, De~Graaff, Davies, Greene, Leja, Brammer, Goulding, Miller, Suess, Weibel, {et~al.}}]{wang2025rubies}
Wang, B., De~Graaff, A., Davies, R.~L., {et~al.} 2025, \href{https://iopscience.iop.org/article/10.3847/1538-4357/adc1ca/meta}{ApJ}, 984, 121

\bibitem[{{Weaver} {et~al.}(2024){Weaver}, {Cutler}, {Pan}, {Whitaker}, {Labb{\'e}}, {Price}, {Bezanson}, {Brammer}, {Marchesini}, {Leja}, {Wang}, {Furtak}, {Zitrin}, {Atek}, {Chemerynska}, {Coe}, {Dayal}, {van Dokkum}, {Feldmann}, {F{\"o}rster Schreiber}, {Franx}, {Fujimoto}, {Fudamoto}, {Glazebrook}, {de Graaff}, {Greene}, {Juneau}, {Kassin}, {Kriek}, {Khullar}, {Maseda}, {Mowla}, {Muzzin}, {Nanayakkara}, {Nelson}, {Oesch}, {Pacifici}, {Papovich}, {Setton}, {Shapley}, {Shipley}, {Smit}, {Stefanon}, {Taylor}, {Weibel}, \& {Williams}}]{2024ApJS..270....7W}
{Weaver}, J.~R., {Cutler}, S.~E., {Pan}, R., {et~al.} 2024, \href{https://iopscience.iop.org/article/10.3847/1538-4365/ad07e0/meta}{ApJ}, 270, 7

\bibitem[{{Wuyts} {et~al.}(2011){Wuyts}, {F{\"o}rster Schreiber}, {Lutz}, {Nordon}, {Berta}, {Altieri}, {Andreani}, {Aussel}, {Bongiovanni}, {Cepa}, {Cimatti}, {Daddi}, {Elbaz}, {Genzel}, {Koekemoer}, {Magnelli}, {Maiolino}, {McGrath}, {P{\'e}rez Garc{\'\i}a}, {Poglitsch}, {Popesso}, {Pozzi}, {Sanchez-Portal}, {Sturm}, {Tacconi}, \& {Valtchanov}}]{2011ApJ...738..106W}
{Wuyts}, S., {F{\"o}rster Schreiber}, N.~M., {Lutz}, D., {et~al.} 2011, \apj, 738, 106, \dodoi{10.1088/0004-637X/738/1/106}

\bibitem[{Yue {et~al.}(2024)Yue, Eilers, Ananna, Panagiotou, Kara, \& Miyaji}]{yue2024stacking}
Yue, M., Eilers, A.-C., Ananna, T.~T., {et~al.} 2024, \href{https://iopscience.iop.org/article/10.3847/2041-8213/ad7eba/meta}{ApJL}, 974, L26

\bibitem[{Zavala {et~al.}(2021)Zavala, Casey, Manning, Aravena, Bethermin, Caputi, Clements, Da~Cunha, Drew, Finkelstein, {et~al.}}]{zavala2021evolution}
Zavala, J., Casey, C., Manning, S., {et~al.} 2021, \href{https://iopscience.iop.org/article/10.3847/1538-4357/abdb27/meta}{ApJ}, 909, 165

\end{thebibliography}

\end{document}